\documentclass[aps,prd,twocolumn,amsmath,amssymb,groupedaddress]{revtex4-1}

\usepackage[utf8]{inputenc}
\usepackage[dvips]{graphicx}
\usepackage{psfrag}
\usepackage{subfigure}

\newcommand{\dd}{\mathrm{d}}
\newcommand{\ee}{\mathrm{e}}
\newcommand{\ii}{\mathrm{i}}

\renewcommand{\vec}[1]{\ensuremath{\boldsymbol{#1}}}

\newcommand{\diag}{\mathrm{diag}}

\usepackage{xspace}
\newcommand*{\eg}{e.\,g.\@\xspace}
\newcommand*{\ie}{i.\,e.\@\xspace}
\newcommand*{\cf}{cf.\@\xspace}
\newcommand*{\eq}[1]{Eq.~(\ref{eq:#1})}
\newcommand*{\fig}[1]{Fig.~\ref{fig:#1}}
\renewcommand*{\sec}[1]{Sec.~\ref{sec:#1}}
\newcommand*{\tab}[1]{Table~\ref{tab:#1}}

\usepackage[ps2pdf=true]{hyperref}
\hypersetup{pdftitle=Structure Formation and Backreaction in Growing
Neutrino Quintessence}

\begin{document}


\title{Structure Formation and Backreaction in Growing Neutrino
Quintessence}

\author{Youness Ayaita}
\author{Maik Weber}
\email[]{m.weber@thphys.uni-heidelberg.de}
\author{Christof Wetterich}
\affiliation{Institut für Theoretische Physik, Universität
Heidelberg\\ Philosophenweg 16, D--69120 Heidelberg, Germany}


\begin{abstract}
	A dependence of the neutrino masses on the dark energy scalar
	field could provide a solution to the {\it why now} problem of
	dark energy. The dynamics of the resulting cosmological model,
	{\it growing neutrino quintessence}, include an attractive force
	between neutrinos substantially stronger than gravity.  We present
	a comprehensive approach towards an understanding of the full
	cosmological evolution including the formation of large--scale
	neutrino structures. Important effects we account for are local
	variations in the dark energy and the backreaction on the
	background evolution, as well as relativistic neutrino velocities.
	For this aim, we develop a relativistic $N$--body treatment of the
	neutrinos combined with an explicit computation of the local
	quintessence field.  At its current stage, the simulation method
	is successful until $z \approx 1$ and reveals a rich
	phenomenology. We obtain a detailed picture of the formation of
	large--scale neutrino structures and their influence on the
	evolution of matter, dark energy, and the late--time expansion of
	the Universe.
\end{abstract}

\pacs{}

\maketitle


\section{Introduction}
\label{sec:introduction}

In today's cosmological standard scenario, a cosmological constant
$\Lambda$ is assumed to explain the observed accelerated expansion of
the Universe. Although so far consistent with all major observational
probes \cite{WoodVasey07, Reid09, Komatsu10, Sullivan11}, it faces two
fundamental and unresolved problems.  Besides its disturbingly tiny
value (the {\it cosmological constant problem}), it remains miraculous
why $\Lambda$ has become important just recently (the {\it why now} or
{\it coincidence problem}).

It is thus tempting to think of alternative cosmological scenarios in
which these two problems are alleviated. In this work, we study
growing neutrino quintessence, which addresses both of the problems.
Growing neutrino quintessence \cite{Amendola07, Wetterich07} relies on
two assumptions. First, the dark energy component is described by a
dynamical scalar field.  Second, the neutrino masses depend on this
field. Studying the background evolution, it is found that the
coincidence problem can indeed be solved for a neutrino--cosmon
coupling somewhat larger than gravity.

In order to explore the implications of this model, its evolution has
to be understood also on the perturbation level. Linear perturbation
theory, however, breaks down even at large scales \cite{Mota08}. This
is due to large overdensities in the neutrino fluid becoming
non--linear at $z \lesssim 2$ on supercluster scales.

An understanding of the non--linear evolution, although of utmost
importance for confronting the model with observational data, is still
lacking. In first attempts, some aspects of the model have been
studied with hydrodynamical and $N$--body methods \cite{Wintergerst09,
Baldi11}. Yet, three crucial aspects of the model have not been
included so far. First, as shown by an analytical study of single
non--linear neutrino overdensities, local variations of the neutrino
mass can become very large \cite{Nunes11}. Second, numerical results
\cite{Baldi11} show that neutrino particles are likely to reach highly
relativistic velocities and thereby leave the Newtonian limit. Third,
the local mass variations as well as the relativistic corrections
change the average energy--momentum tensor of neutrinos; this {\it
backreaction} effect alters the background evolution of the dark
energy scalar field.

It is the purpose of this work to form the basis for an adequate
$N$--body based simulation method capable of including all major
effects of growing neutrino quintessence. For the first time, we
implement an explicit computation of the local dark energy scalar
field for every time step.  This allows us to account for local mass
variations and backreaction effects. We present a relativistic
treatment of the neutrino dynamics as appropriate for the high
velocities of neutrinos occurring during the structure formation
process. The results of linear perturbation theory are used for 
the initial conditions. All these aspects are not merely
small corrections, but decisive for every quantitative prediction.

The organization of this paper is as follows. We present growing
neutrino quintessence and derive the equations of motion required for
our simulation in \sec{growing}. In \sec{strategy}, we discuss
strategies how to account for the peculiar features of growing
neutrino quintessence in a numerical simulation. The numerical results
are presented in Secs.~\ref{sec:neutrino} and \ref{sec:impact}. At
first, we have a look at the formation of neutrino structures
(\sec{neutrino}), and then explore the effects on the evolution of
dark energy and matter (\sec{impact}). We summarize in \sec{summary}.

\section{Growing Neutrino Quintessence}
\label{sec:growing}

\subsection{Overview}
\label{sec:overview}

Growing neutrino quintessence is a possible solution to both the {\it
cosmological constant} and the {\it why now} problem of dark energy
\cite{Amendola07, Wetterich07}. Since it is a quintessence model with
a dark energy scalar field, the cosmon $\varphi$, dark energy evolves
dynamically.  During most of the cosmological evolution, its energy
density decays similarly to the densities of the other species. The
vanishing of the cosmon potential $V(\varphi)$ for large values of
$\varphi$ can be rooted in the approach to a fixed point with
effective dilatation symmetry. The scaling solution with dark energy
of the same order of magnitude as radiation or smaller can then
explain the tiny overall size of the dark energy density by the large
age of the Universe. 

In contrast to standard quintessence scenarios, however, a coupling
between the cosmon $\varphi$ and the neutrinos provides a solution to
the coincidence problem as well. Due to their small masses,
cosmological neutrinos have become non--relativistic only recently.
This event triggers the onset of dark--energy domination in growing
neutrino quintessence.

A coupling between the cosmon $\varphi$ and a matter species, here
assumed to be the neutrinos, is expressed by the exchange of energy
and momentum. The individual energy--momentum tensors do not satisfy
a conservation equation, only their sum does. Denoting with
$T^{\alpha\beta}$ the energy--momentum tensor of the neutrinos and
with $S^{\alpha\beta}$ the energy--momentum tensor of the cosmon
field, we have
\begin{equation}
	\nabla_\beta T^{\alpha\beta} = Q^\alpha,\
	\nabla_\beta S^{\alpha\beta} = -Q^\alpha.\
	\label{eq:exchange}
\end{equation}
Since no known symmetry requires $Q = 0$, we generally have to expect
a non--vanishing coupling. A specific form of the coupling proposed by
early works \cite{Wetterich94, Amendola99} is
\begin{equation}
	Q^\alpha =  -\beta \,T \, \partial^\alpha \varphi
	\label{eq:coupling}
\end{equation}
with a dimensionless coupling parameter $\beta$ and $T \equiv
{T^\alpha}_\alpha$, using units where the reduced Planck mass $M_P = (8 \pi
G)^{-1/2}$ is set to unity. 
Writing $T = -\rho_\nu + 3 p_\nu$, we see that $T$ vanishes as long as
the neutrinos are relativistic, $w_\nu = p_\nu/ \rho_\nu = 1/3$. The
coupling only becomes effective once the neutrinos have become
non--relativistic. For large values of the coupling $\beta$, the
coupling can stop the evolution of the cosmon. An almost constant
value of the cosmon potential leads then to an onset of dark--energy
domination at recent times, similar to the concordance model $\Lambda$CDM. 

On a particle physics level, the coupling, \eq{coupling}, is realized
as a dependence of the neutrino masses on the cosmon field.  The
coupling parameter $\beta$, in terms of the average neutrino mass
$m_\nu \equiv m_\nu(\varphi)$ \cite{Amendola07}, now is 
\begin{equation}
	\beta = - \frac{\dd \ln m_\nu }{ \dd \varphi}.
	\label{eq:betadef}
\end{equation}
We further assume that the coupling parameter $\beta$ is constant,
\ie, it does not depend on $\varphi$.  In this case, we have the
crucial relation
\begin{equation}
	m_\nu(\varphi) \propto \ee^{-\beta\varphi}.
	\label{eq:mnu}
\end{equation}
In scenarios with an expansion history similar to $\Lambda$CDM, the
coupling $\beta$ takes large negative values, typically of order
$\beta \sim -10^2$.  The possible couplings to other matter species
are assumed to be negligible. The self--interaction of the cosmon is
given by a potential $V(\varphi)$, for which an example is the
exponential potential \cite{Amendola07},
\begin{equation}
	V(\varphi) \propto \ee^{-\alpha \varphi},
	\label{eq:vofphi}
\end{equation}
where $\alpha$ is a dimensionless model parameter with typical values
$\alpha \gtrsim 10$ to satisfy early dark energy constraints
\cite{Doran07, Reichardt11}.

The coupling, strong enough to stop the cosmon evolution in the
background, also crucially modifies the evolution of perturbations. In
fact, if the neutrinos are non--relativistic, the cosmon perturbation
$\delta \varphi$ is approximately a factor of $2 \beta$ larger than
the neutrino--induced gravitational potential. The perturbation
$\delta \varphi$ describes an attractive force between the neutrinos
of order $|\vec F| \approx |\beta \vec \nabla \delta \varphi| \approx
2 \beta^2 |\vec F_\mathrm{gravity}|$ \cite{Wintergerst09}. Since
realistic scenarios have $\beta^2 \gg 1$, the extra force leads to a
very rapid growth of perturbations in the neutrino fluid becoming
non--linear at $z_\mathrm{nl} \approx 1$--$2$, even on large scales
\cite{Mota08}.

Analyzing growing neutrino quintessence quantitatively thus requires
adequate methods to study the non--linear evolution. Perturbations
$\delta \varphi$ in the quintessence field imply local mass variations
$\delta m_\nu$ by virtue of \eq{mnu}. These variations can
significantly change the averaged energy--momentum tensor entering the
background equations. This backreaction effect of structure formation
on the background evolution will turn out to be crucial. Furthermore,
the forces during the non--linear evolution are strong enough to
accelerate the neutrinos to relativistic velocities. We thus have to
work with a relativistic description. For these aims, we shall next
derive and collect the necessary equations.

\subsection{Fundamental equations}
\label{sec:fundamental}

In a first step, we will present a fundamental definition of the model
in terms of an action and derive the basic equations,
Eqs.~(\ref{eq:exchange}) and (\ref{eq:coupling}), introduced above. In
principle, one has to describe all three neutrino species together.
This can be done in growing neutrino quintessence \cite{Wetterich07}.
In what follows, however, we adopt the simplified working hypothesis
of the neutrinos having all the same mass. The three flavors of
neutrinos enter then only in the initial number density of neutrinos.
For our simulation, we can deal effectively with one species. 

The dynamics of the cosmon field $\varphi$ and the neutrino field
$\psi$ are described by a usual scalar field Lagrangian ${\mathcal
L}_\varphi$ and a Majorana Lagrangian ${\mathcal L}_\nu$ respectively,
with the peculiarity that the neutrino mass term $m_\nu \, \bar \psi
\psi$ is assumed to depend on $\varphi$,
\begin{align}
	{\mathcal L}_\varphi &= - \frac{1}{2} \partial_\alpha \varphi \,
	\partial^\alpha \varphi - V(\varphi),
	\label{eq:lphi}
	\\
	{\mathcal L}_\nu &= \ii\,\bar \psi \, \left( \gamma^\alpha
	\nabla_\alpha +
	m_\nu(\varphi) \right)\, \psi.
	\label{eq:lnu}
\end{align}
Here, we use the vierbein formalism to describe the neutrino field in
curved spacetime \cite{Brill57}. The quantities $\gamma^\alpha(x)$ are
related to the usual Dirac matrices $\gamma^a$ ($a = 0,1,2,3$) by
virtue of the vierbein $e^\alpha_a(x)$: $\gamma^\alpha(x) = \gamma^a
e^\alpha_a (x)$, where the vierbein is related to the metric by
$g^{\alpha\beta} = e^\alpha_a e^\beta_b \eta^{ab}$, $\eta^{ab}=\diag
(-1,1,1,1)$. The action reads
\begin{equation}
	S = \int_{}^{} \dd^4 x\, \sqrt{-g}\, \left( \mathcal L_0 +
	\mathcal L_\nu + \mathcal L_\varphi \right),
	\label{eq:fullaction}
\end{equation}
where $\mathcal L_0$ includes the remaining cosmological species and
gravity. A Majorana constraint relates $\bar \psi$ to $\psi$. 

We obtain the energy--momentum tensor ${T}^{\alpha\beta}$
for the neutrino field by the usual definition
\begin{equation}
	\delta S_\nu = \frac{1}{2} \int_{}^{} \dd^4 x \, \sqrt{-g}\,
	(\delta g_{\alpha\beta}) \, {T}^{\alpha\beta}.
	\label{}
\end{equation}
We assume a diagonal metric, which we have, \eg, in the Newtonian
gauge---otherwise, one employs a generalization with vierbeins. The
energy--momentum tensor reads
\begin{equation}
	{T}^{\alpha\beta} = -\frac{\ii }{2} \, \bar \psi
	\gamma^{(\beta} \nabla^{\alpha)} \psi + \frac{\ii}{2}
	\nabla^{(\alpha}
	\bar \psi \gamma^{\beta)} \psi,
	\label{eq:nutalphabeta}
\end{equation}
where we have symmetrized in the indices $\alpha$ and $\beta$ (\cf,
\eg, \cite{Brill57}). The equations of motion following from
${\mathcal L}_\nu$ are the Dirac equations in curved spacetime,
\begin{align}
	\gamma^\alpha \nabla_\alpha \psi + m_\nu(\varphi) \psi &=0,
	\nonumber
	\\
	-\nabla_\alpha \bar \psi \gamma^\alpha + m_\nu(\varphi) \bar \psi &=0.
	\label{eq:dirac2}
\end{align}
In the absence of the coupling, \ie in the case $m_\nu =
\text{const.}$, these equations imply the energy--momentum
conservation equation $\nabla_\beta {T}^{\alpha\beta} = 0$ for the
neutrino field.

In growing neutrino quintessence, however, the mass depends on the
spacetime coordinates $x$ via $\varphi(x)$, and the derivative also
acts on $m_\nu$. Inserting the equations of motion into $\nabla_\beta
{T}^{\alpha\beta}$, we find
\begin{equation}
	\nabla_\beta {T}^{\alpha\beta} =\partial^\alpha m_\nu(\varphi)\,
	\ii \bar \psi \psi
	\equiv -\partial^\alpha m_\nu(\varphi)\, \tilde
	n_\nu.	
	\label{}
\end{equation}
We will give an interpretation of $\tilde n \equiv -\ii \bar \psi
\psi$ below. Using the definition of $\beta$, \eq{betadef}, we may
write
$\partial^\alpha m_\nu(\varphi) = - \beta m_\nu(\varphi)
\partial^\alpha \varphi$. In our discussion, an important role is
played by the trace $T$ of the neutrino energy--momentum tensor,
\begin{equation}
	T \equiv {T^\alpha}_\alpha = -m_\nu(\varphi)\, \tilde n_\nu =
	-\rho_\nu + 3 p_\nu.
	\label{eq:trtdef}
\end{equation}
This eventually shows the form of the energy--momentum exchange, \cf
\eq{coupling},
\begin{equation}
	\nabla_\beta {T}^{\alpha\beta} = - \beta\, T 
	\, \partial^\alpha \varphi.
	\label{eq:consnu}
\end{equation}
A standard computation for the energy--momentum tensor  of the cosmon,
\begin{equation}
	S^{\alpha\beta} = \partial^\alpha \varphi\, \partial^\beta
	\varphi + g^{\alpha\beta} \mathcal L_\varphi,
	\label{eq:tphi}
\end{equation}
leads to 
\begin{equation}
	\nabla_\beta {S}^{\alpha\beta} = + \beta\, T \, 
	\partial^\alpha \varphi,
	\label{eq:consphi}
\end{equation}
and one verifies the conservation equation $\nabla_\beta \left(
{T}^{\alpha\beta} + {S}^{\alpha\beta} \right) = 0$ for the sum of the
neutrino and cosmon energy--momentum tensors.

In the non--relativistic limit, $\tilde n_\nu$ introduced above
corresponds to the neutrino number density $n_\nu$. In contrast to
$\tilde n_\nu$, however, $n_\nu$ does not transform as a scalar. For a
Lorentz transformation, \eg, $n_\nu$ picks up the volume contraction
factor $1/\gamma$. This will become more concrete when we represent
the neutrino field by effective relativistic particles (see
\sec{particles}).

\subsection{Cosmon dynamics}
\label{sec:cosmon}

The coupling between neutrinos and the cosmon field has important
impacts on the evolution of both species. Variation of the action
with respect to $\varphi$ yields the modified Klein--Gordon equation
\begin{equation}
	\nabla^\alpha \nabla_\alpha \varphi - V_{,\varphi}(\varphi) =
	 \beta \, T.
	\label{eq:kgmod}
\end{equation}
We split into background quantities (spatial averages only depending
on time) and perturbations (spatially varying with vanishing mean),
$g_{\alpha\beta} = \bar g_{\alpha\beta}+\delta g_{\alpha\beta},
\varphi = \bar \varphi + \delta \varphi$, and $T^{\alpha\beta} = {\bar
T}^{\alpha\beta} + \delta T^{\alpha\beta}$. We assume that the metric
perturbations $\delta g_{\alpha\beta}$ and the cosmon perturbation
$\delta \varphi$ can be treated in linear approximation and that their
time derivatives are small.

We choose the conformal Newtonian gauge (see, \eg, \cite{Ma95}), in
which the metric reads
\begin{equation} 
	\dd s^2 = a^2 \left(- (1+2\Psi) \dd \eta^2 + (1-2\Phi) \dd \vec x^2
	\right), 
	\label{eq:metric}
\end{equation}
with the conformal time $\eta$, the comoving coordinates $\vec x$, the
scale--factor $a(\eta)$, and the two scalar potentials $\Psi$ and
$\Phi$.  It is now straightforward to evaluate \eq{kgmod}. Separating
into a background part, independent of the spatial position, and a
linear perturbation part, we obtain two evolution equations.

For the background quantities, we find
\begin{equation}
	\bar \varphi'' + 2 {\mathcal H} \bar \varphi' + a^2
	V_{,\varphi}(\bar \varphi) =  a^2 \beta\, (\bar \rho_\nu - 3 \bar
	p_\nu),  
	\label{eq:bgphi}
\end{equation}
where primes denote derivatives with respect to $\eta$, $\mathcal H
\equiv a'/a$ is the conformal Hubble parameter, and we have used $\bar
T = -\bar \rho_\nu + 3 \bar p_\nu$. Obviously, the coupling is
ineffective as long as $w_\nu = \bar p_\nu/\bar \rho_\nu = 1/3$, \ie
if the neutrinos are relativistic.  Once the neutrinos turn
non--relativistic, the right--hand side of \eq{bgphi} stops the
further evolution of $\bar \varphi$. By this mechanism, the model can
solve the coincidence problem \cite{Amendola07, Wetterich07}. 

In the perturbations, the equation reads
\begin{align} 
	\Delta \delta \varphi - 
	a^2	V_{,\varphi \varphi}(\bar \varphi) \delta \varphi
	&+ 2 \Psi (\bar \varphi'' + 2 {\mathcal H} \bar \varphi') =
	\nonumber\\
	&=  a^2 \beta\, \delta T,
	\label{eq:pertphi}
\end{align}
where the spatial derivatives refer to comoving coordinates.  In the
fluid description, $\delta T = -\delta\rho_\nu + 3 \delta p_\nu$. For
our purpose, however, it is more convenient to calculate $\delta T$
directly from the distribution of particles in our simulation.

\subsection{Neutrino dynamics}
\label{sec:neutrinodynamics}

In this section, we investigate the motion of a neutrino particle with
a cosmon--depending mass $m_\nu(\varphi)$.  We describe a neutrino as
a classical, yet relativistic particle with world line
$\xi^\alpha(\tau)$ and four--velocity $u^\alpha=\dd \xi^\alpha/\dd
\tau$, where $\tau$ denotes the particle's proper time, defined via
$\dd \tau^2 = -g_{\alpha\beta}\dd \xi^\alpha \dd \xi^\beta$.  The
energy--momentum tensor of this particle is given by
\begin{equation}
	T^{\alpha\beta} = \frac{1}{\sqrt{-g}}\int
	\dd\tau\,m_\nu(\varphi(\xi))\, u^\alpha
	u^\beta \delta^4(x-\xi), 
	\label{eq:tparticle}
\end{equation}
where $g$ is the determinant of the metric and $\delta^4(x)$ denotes
the four--dimensional Dirac delta function. The factor $1/\sqrt{-g}$
ensures the correct normalization of the Dirac delta function in
curved spacetime by compensating the invariant volume form
$\sqrt{-g}\,\dd^4 x$.
The one--particle action is constructed from the energy--momentum
tensor,
\begin{equation}
	S_\nu = \int_{}^{} \dd^4 x \, \sqrt{-g}\ T^{\alpha\beta}
	g_{\alpha\beta} = - \int_{}^{} \dd \tau \, m_\nu(\varphi(\xi)).
	\label{eq:sparticle}
\end{equation}
The equations of motion can be obtained by varying $S_\nu$ with
respect to the particle's path $\xi(\tau)$. In the uncoupled case,
$m_\nu = \text{const.}$, this would give the standard geodesic
equation. The modifications due to the cosmon--neutrino coupling are
the same that we will find below by using energy--momentum
conservation.

In what follows, we will use \eq{tparticle} in the energy--momentum
conservation equation, \eq{consnu}, in order to derive the equations
of motion.  Modifications to the standard geodesic equation enter in
two ways, on the right--hand side through the exchange term
$Q^\alpha=-\beta\, T \,\partial^\alpha \varphi$ and on the left--hand
side through the cosmon--depending mass $m_\nu(\varphi)$.

Since $u^\alpha u_\alpha = -1$, the right--hand side simply becomes
\begin{equation}
	-\beta \, T \, \partial^\alpha \varphi =
	\frac{1}{\sqrt{-g}}\int_{}^{}\dd\tau\, m_\nu(\varphi) \,\beta\,
	\partial^\alpha\varphi\, 
	\delta^4(x-\xi).
	\label{eq:rhsgeo}
\end{equation}
The left--hand side is
\begin{equation}
	\nabla_\beta T^{\alpha\beta} = \partial_\beta T^{\alpha\beta} +
	\Gamma^\alpha_{\beta\lambda}T^{\lambda\beta} +
	\Gamma^\lambda_{\beta\lambda}T^{\alpha\beta}.
	\label{eq:covderiv}
\end{equation}
It is straightforward to show
\begin{equation}
	\partial_\beta \left(\sqrt{-g}\, T^{\alpha\beta}\right) =
	\int_{}^{}\dd \tau\, \frac{\partial}{\partial\tau}\left(
	m_\nu(\varphi) u^\alpha \right) \delta^4(x-\xi).
	\label{eq:delt}
\end{equation}
The derivative acting on $m_\nu(\varphi)$ is $\partial
m_\nu(\varphi)/\partial \tau = -\beta\, m(\varphi)\,u^\lambda
\partial_\lambda \varphi$. Apart from this contribution, \ie in the
uncoupled case, \eq{covderiv} reproduces the standard geodesic
equation. Thus, we find
\begin{align}
	\nabla_\beta T^{\alpha\beta} = &\frac{1}{\sqrt{-g}} \int \dd
	\tau\, m_\nu(\varphi)\,\delta^4(x-\xi) \nonumber \\ & \times
	\left( \frac{\dd u^\alpha}{\dd \tau} +
	\Gamma^\alpha_{\rho\sigma}u^\rho u^\sigma - \beta \,u^\lambda
	\partial_\lambda \varphi\, u^\alpha \right).  \label{eq:lhsgeo}
\end{align}
Comparing this with \eq{rhsgeo}, we arrive at the equation of motion:
\begin{equation}
	\frac{\dd u^\alpha}{\dd \tau} + \Gamma^\alpha_{\rho\sigma}u^\rho
	u^\sigma = \beta \,\partial^\alpha \varphi + \beta\, u^\lambda
	\partial_\lambda \varphi\, u^\alpha, \label{eq:eomnu}
\end{equation}
which describes the deviation from the geodesic motion due to the
cosmon--neutrino coupling. This equation of motion can also be
obtained by a conformal transformation of the geodesic equation
\cite{Baldi11}. Let us give a brief interpretation of its terms.
\begin{itemize}
	\item $\Gamma^\alpha_{\rho\sigma}u^\rho u^\sigma$ describes the
		gravitational effects. In the component $\alpha=0$, this is
		the Hubble damping. The spatial components give rise to the
		gravitational force. In the Newtonian limit, this would be
		given by $\vec \nabla \Psi$. In the relativistic case, the
		situation is more complicated involving both potentials,
		$\Psi$ and $\Phi$.
	\item $\beta\, \partial^\alpha \varphi$ is the cosmon--mediated
		{\it fifth force}. In the Newtonian limit, this corresponds to an
		attractive force between neutrinos about $2 \beta^2$ times stronger
		than gravity \cite{Wintergerst09}. 
		
	\item $\beta\, u^\lambda \partial_\lambda \varphi \, u^\alpha$
		represents a velocity dependent force. It can be understood as
		a consequence of momentum conservation. Particles are
		accelerated when they move into a direction where they lose
		mass. It also modifies the universal damping term, which is no longer
		proportional to the Hubble parameter $\mathcal H$, but to $(\mathcal H - \beta
		\varphi')$.
\end{itemize}
It is instructive to consider the case of a static particle, $u^i=0$.
Then, the time component of the right--hand side of \eq{eomnu}
vanishes, and the spatial components reduce to $\beta\, \partial^i
\varphi$. Thus, a static particle is only affected by the spatial
variation $\vec \nabla \varphi$ and not by $\varphi'$, and $u^0 =
(1-\Psi)/a$ does not depend on $\varphi$.

\section{Strategy and method}
\label{sec:strategy}

Although many efforts to describe the structure formation process in
growing neutrino quintessence have been made (\cf, \eg,
\cite{Brouzakis07, Mota08, Wintergerst09, Baldi11}), none of the
applied methods was capable of drawing a comprehensive picture, nor of
providing reliable quantitative results once non--linearities are
strong.

These challenges motivate the development of a new method,
specifically designed for growing neutrino quintessence. We describe
the framework in \sec{framework}. As a fundamental ingredient, we keep
track of the perturbed cosmon scalar field allowing us to incorporate
local mass differences of the neutrinos. We find a significant
neutrino mass suppression once a large fraction of the neutrinos is
bound in non--linear structures.  This induces a {\it backreaction} on
the evolution of the cosmological background (\sec{backreaction}). The
initial conditions are drawn from the linear results (\sec{initial}).
We discuss numerical issues in \sec{numerical}. 

\subsection{Framework}
\label{sec:framework}

An important goal is to make predictions for the density contrasts
$\delta_\nu(x)$, $\delta_m(x)$ and the peculiar velocity fields $\vec
v^\text{pec}_\nu(x)$, $\vec v^\text{pec}_m(x)$ of neutrinos and matter respectively. These
are linked to the gravitational potential and to various observables.
These fields, however, do not carry all the necessary information
needed to describe their evolution. They merely arise from the first
moments of the full phase--space distribution functions $f_\nu(\eta,
x^i, v_j)$, $f_m(\eta, x^i, v_j)$. In a non--linear evolution, higher 
moments of these distributions become important. It is thus most
efficient to directly sample the distribution functions by a finite
number of effective particles, $N_\nu$ and $N_m$ respectively. These
particles carry a comoving position $\vec x$, a velocity $\vec 
v = \dd \vec x / \dd \eta$, and a rest mass $M_\nu$ (neutrinos), or
$M_m$ (matter).  Our method thus bases upon an $N$--body scheme.

The motions of these particles depend on the cosmological background,
the two gravitational potentials $\Psi(x)$, $\Phi(x)$, and the cosmon
$\varphi(x) = \bar \varphi(\eta) + \delta \varphi(x)$. In the
Newtonian limit and neglecting the local mass variation of neutrinos,
both gravity and the fifth force can be described as two--body forces.
If we want to go beyond these approximations, we need to know the
explicit values of these fields.

We model the fields $\Psi$, $\Phi$, and $\delta \varphi$ as discrete
values on a three--dimensional grid in a volume $V$ with $N_c$ cells.
The volume is a cube with side length $L$, every cell has the volume
$(\Delta x)^3 = V / N_c$. We employ periodic boundary conditions. For
the concrete values chosen, \cf \tab{values}.
\begin{table}[htb!]
	\begin{center}
		\begin{tabular}{l|c}
			{\it Simulation properties} & {\it Specification} \\ \hline
			Box volume $V = L^3$ & $600^3\, h^{-3}\text{Mpc}^3$ \\
			Number of cells $N_c$ & $256^3$ \\ \hline
			Neutrino particles $N_\nu$ & $2\times 10^7$ \\ 
			Matter particles $N_m$ & $2\times 10^7$ \\ \hline
			Initial redshift $z_i$ & $4$ (neutrinos), $49$ (matter)\\ 
			Final redshift $z_f$ & $1$ \\ \hline
			Particle properties & $\vec x$, $\vec v$, $M_\nu$, $M_m$ \\
			Dynamical fields & $\Psi$, $\Phi$, $\delta \varphi$ \\
			\hline
			Background quantities & $\mathcal H$, $\bar \varphi$,
			$\bar{\rho}_\nu$, $\bar{p}_\nu$, $\bar{\rho}_m$
		\end{tabular}
		\caption{The basic parameters and ingredients of the
		simulation.}
		\label{tab:values}
	\end{center}
\end{table}

Let us outline the basic steps of the algorithm we use. The details
will be given in subsequent sections.
\begin{itemize}
	\item[(1)] Initialization of the simulation (\cf \sec{initial}).
		\begin{itemize}
			\item[(a)] Generate a realization of initial fields
				$\delta_\nu$, $\delta_m$, $\vec v^\text{pec}_\nu$, $\vec
				v^\text{pec}_m$ from the linear spectra.
			\item[(b)] Sample the corresponding distribution functions
				with $N_\nu$ effective neutrino and $N_m$ effective
				matter particles.
		\end{itemize}
	\item[(2)] Simulation steps.
		\begin{itemize}
			\item[(a)] Accelerate and move the effective neutrino and
				matter particles (\cf Secs.~\ref{sec:particles} and
				\ref{sec:neutrinodynamics}).
			\item[(b)] Calculate the potentials $\Psi$, $\Phi$, and
				the cosmon perturbation $\delta\varphi$ (\cf
				\sec{particles}). Update neutrino masses $M_\nu$ with
				the new local cosmon field $\varphi$ (\cf
				\sec{backreaction}).
			\item[(c)] Measure the averages ${\bar \rho}_\nu$, ${\bar
				p}_\nu$. With these quantities, evolve the background
				cosmon $\bar \varphi$. The Hubble parameter $\mathcal
				H$ is evaluated by Friedmann's equation $3 \mathcal
				H^2 / a^2 = {\bar \rho}_\nu + {\bar \rho}_m + {\bar
				\rho}_\varphi$ (\cf \sec{backreaction}).
		\end{itemize}
\end{itemize}

\subsection{Particles and fields}
\label{sec:particles}

As stated above, we assume that the fields $\Psi$, $\Phi$, and $\delta
\varphi$ can be treated linearly and that their time derivatives,
$\Psi'$, $\Phi'$, and $\delta \varphi'$, are small. This leads to
simple algebraic equations in Fourier space that allow us to calculate
the fields from a given distribution of particles. This method
requires to carry out several Fourier transforms at each time step,
which can be done efficiently by the use of Fast Fourier Transform
routines. Let us collect the relevant equations.

The gravitational potential $\Phi$, on subhorizon scales, is obtained
from the Poisson equation \cite{Ma95},
\begin{equation}
	k^2 \Phi = \frac{a^2}{2} \delta \rho, 
	\label{eq:poisson}
\end{equation}
where $k = |\vec k|$ and $\delta\rho = \delta \rho_\nu + \delta \rho_m
+ \delta \rho_\varphi$. 

While we can obtain $\delta \rho_\varphi$ from the cosmon field and its
perturbation,
\begin{equation}
	\delta \rho_\varphi = \frac{\bar \varphi'\,
	\delta\varphi}{a^2}+V(\bar \varphi)\, \delta\varphi,
	\label{eq:drhophi}
\end{equation}
we need to calculate $\delta\rho_\nu$ (and $\delta\rho_m$ analogously)
from the effective particles.  

Since $\delta \rho_\nu = - \delta {T^0}_0$, we get the contribution of
one particle at position $\vec \xi$ from \eq{tparticle},
\begin{equation}
	-{T^0}_0 = \frac{1}{\sqrt{\tilde g}}\, \gamma\, M_\nu\,
	\delta^3(\vec x- \vec \xi) ,
	\label{eq:t00nu}
\end{equation}
where we have introduced the Lorentz factor,
\begin{equation}
	\gamma \equiv \frac{\sqrt{- g_{00}}\dd x^0}{\dd \tau} =
	\frac{1}{\sqrt{1-(1-2\Psi-2\Phi)\vec v^2}},
	\label{eq:gamma}
\end{equation}
and the determinant of the spatial metric, 
\begin{equation}
\tilde g = \det ( g_{ij} ),\ \ 	\sqrt{\tilde g} = a^3\,(1-3\Phi).
\label{eq:detgtilde}
\end{equation}
In the case of matter, we can approximate $\vec v^2 \ll 1$. In
contrast, we keep the full relativistic equations for neutrinos since
values of $\vec v^2$ close to one can be reached once large neutrino
structures form. 

Finally, the discrete field value $\rho_\nu(\vec x)$ at a cell $\vec
x$ is obtained by summing up the contributions of all the particles
located inside the cell. Subtracting the mean $\bar \rho_\nu$ yields
$\delta\rho_\nu(\vec x)$. The Poisson equation, \eq{poisson}, can now
be solved as follows. The sum $\delta \rho(\vec x) = \delta \rho_\nu
(\vec x) + \delta \rho_m (\vec x) + \delta \rho_\varphi (\vec x)$ is
Fourier transformed and inserted into \eq{poisson}. The resulting
field $\Phi(\vec k)$ and its gradient $\ii \vec k \Phi(\vec k)$ are
then transformed back to real space, yielding $\Phi(\vec x)$ and $\vec
\nabla \Phi(\vec x)$.  

The second gravitational potential $\Psi$ equals $\Phi$ as long as
anisotropic stress is negligible, \ie when $ {T^i}_j$ is small for $i
\not= j$. Inspecting the energy--momentum tensor of a single particle,
\cf \eq{tparticle}, the off--diagonal components ${T^i}_j$ are of
order $\vec v^2$ and thus only important for relativistic species. In
our case, the only contribution comes from the neutrinos. Perturbation
theory yields the following expression for the difference of the two
potentials \cite{Ma95}:
\begin{equation}
	k^2 (\Phi - \Psi) = -\frac{3}{2}\,a^2\, \sum_{i,j}\left(\frac{
	k_i k_j}{k^2}-\frac{1}{3}\delta_{ij}\right)
	{\Sigma^i}_j, 
	\label{eq:phimpsi}
\end{equation}
with the traceless components of the neutrino energy--momentum tensor, 
\begin{equation}
	{\Sigma^i}_j = {T^i}_j -
	\frac{1}{3}\delta^i_j\, {T^k}_k.
	\label{eq:sigmaij}
\end{equation}
The field ${\Sigma^i}_j(x)$ is obtained from the neutrino
particles by using the form of the one--particle energy--momentum
tensor, \eq{tparticle}.  A straightforward calculation yields
\begin{equation} 
	{\Sigma^i}_j =  \frac{1-2\Psi-2\Phi}{\sqrt{\tilde
	g}} \gamma M_\nu \left( v^i v_j - \frac{\vec v^2}{3} \right)
	\delta^3(\vec x - \vec \xi).
	\label{eq:sigmaijpart}
\end{equation}

The third field, $\delta \varphi$, is calculated from \eq{pertphi} in
Fourier space. On the right--hand side, we need $\delta T = \delta
{T^\alpha}_\alpha$. Again from \eq{tparticle}, the one--particle
contribution is
\begin{equation}
	T = -\frac{1}{\sqrt{\tilde g}}\, \frac{M_\nu}{\gamma}\,
	\delta^3(\vec x - \vec \xi).
	\label{eq:trtpart}
\end{equation}

Given gravitational potentials $\Psi$, $\Phi$, and the cosmon
perturbation $\delta\varphi$, the acceleration of the effective
particles is obtained from the corresponding equations of motion. In
the case of neutrinos, this is \eq{eomnu}. In the case of matter, the
right--hand side vanishes, and the equation specializes to the usual
geodesic equation.

For matter particles, we make the approximation $\vec v^2 \ll 1$, and
simply find
\begin{equation}
	\frac{\dd \vec v}{\dd \eta} = - \vec \nabla \Psi - \mathcal H \,
	\vec v.
	\label{eq:eomm}
\end{equation}
For the neutrinos, it is adequate to solve the equation of motion,
\eq{eomnu}, directly for the spatial components of the four--velocity
$u^i = \gamma v^i (1-\Psi)/a$. Since this automatically respects
$u^\alpha u_\alpha = -1$, the speed of light limit is robustly
enforced. A differential equation for $\vec v$ itself could be
unstable in this regard.

The gradients $\vec \nabla \Psi$, $\vec \nabla \Phi$, $\vec \nabla
\delta \varphi$, occurring in the equations of motion are calculated
in Fourier space and then transformed back to position space like the
fields $\Psi$, $\Phi$, $\delta\varphi$ themselves. This limits the
accuracy on scales comparable to the cell size $\Delta x$. In
particular, forces between two particles located in the same cell are
completely neglected. We will investigate the impact of this simplification in
\sec{numerical}.

Fourier transformations are also used to estimate the power spectrum of
perturbation variables, \eg of the neutrino density contrast $\delta_\nu(\vec
x)$. The spectrum $P_\nu$ follows from the definition,
\begin{equation}
	\left< \delta_\nu(\vec k)\, \delta_\nu^*(\vec q)\right> = (2
	\pi)^3\,P_\nu(k)\, \delta^3(\vec k - \vec q),	
	\label{eq:powerestimate}
\end{equation}
which we discretize on the grid.

\subsection{Backreaction effects}
\label{sec:backreaction}

The usual way to study cosmological dynamics is to perform a complete
split between the background and the perturbation evolution. First,
averaged energy--momentum tensors and the unperturbed Friedmann metric
are used to find the evolution of the background quantities. Second,
the evolution of perturbations on the pre--calculated background is
investigated.

This procedure neglects the modifications to the background evolution
due to the presence of perturbations, the so--called {\it
backreaction}. In the standard $\Lambda$CDM case, this is a reasonable
approximation since gravitational backreaction is small
\cite{Wetterich01}.  In growing neutrino quintessence, in stark
contrast, the backreaction plays a decisive role. For coupled
quintessence models with couplings of super--gravitational strength, a
significant impact of the non--linear structure formation on the
background evolution is expected \cite{Wetterich01, Schrempp09,
Nunes11, Baldi11}. 

We now illustrate the importance of backreaction in our specific model
and explain how we account for it in the simulation.  The background
dynamics of growing neutrino quintessence differ from $\Lambda$CDM
because they include the joint evolution of the cosmon $\bar \varphi$
and of the neutrinos described by their energy--momentum tensor ${\bar
T}^\alpha_{\ \beta}$. The equation describing this evolution is the
modified Klein--Gordon equation, \eq{bgphi}. Its right--hand side is
given by the quantity $\bar T = {\bar T}^\alpha_{\ \alpha} = -\bar
\rho_\nu + 3 \bar p_\nu$.

In the standard procedure, neglecting backreaction, one would estimate
${\bar T}$ from the evolution of the averaged energy--momentum tensor.
Once the neutrinos have become non--relativistic, the trace is simply
given by the average energy density, ${\bar T} = - \bar \rho_\nu$,
whose evolution follows, \cf \cite{Mota08},
\begin{equation}
	\bar \rho_\nu' + 3 \mathcal H \bar \rho_\nu = - \beta \bar
	\varphi' \bar \rho_\nu.
	\label{eq:trtbackground}
\end{equation}
This equation tells us that ${\bar T}$ will depend on the evolution of
the averaged cosmon $\bar \varphi$ but does not take into account
local variations $\delta \varphi$.

These local perturbations, however, are very important when we regard
the evolution of neutrino structures. A large part of the neutrinos is
concentrated in non--linear structures where $\delta \varphi$ takes
negative values. This leads to a systematic suppression of the masses
$M_\nu \propto \exp(-\beta \delta\varphi)$.  We shall see that
\eq{trtbackground} thus leads to a wrong estimate of ${\bar T}$.

The correct method to estimate ${\bar T}$ taking backreaction into
account is to use the exact expression, which we obtain from the
one--particle contributions, \eq{trtpart}. Performing the spatial
averaging, we get as a sum over the particles $p$,
\begin{equation}
	{\bar T} = \frac{\int_{V}^{} \dd^3 x\, \sqrt{\tilde g}\,
	T}{\int_{V}^{} \dd^3 x\, \sqrt{\tilde g}} = \frac{-1}{a^3 V}
	\sum_{p}^{} \frac{M_\nu(\varphi(\vec \xi_p))}{\gamma_p}.
	\label{eq:trttrue}
\end{equation}
This expression underlines the influence of the perturbations. The
quantity ${\bar T}$ is suppressed by two essential effects. First, we
clearly see the dependence on the particles' actual masses determined
by the local cosmon field. Second, once the fifth force has
accelerated the particles to relativistic velocities, the Lorentz
factors $\gamma>1$ lead to an additional suppression. 

\begin{figure}[htb!]
	\begin{center}
		\psfrag{xlabel}[B][c][.8][0]{redshift $z$}
		\psfrag{ylabel}[B][c][.8][0]{$-{\bar T}$, $\bar \rho_\nu$
		[$10^{-8}/\text{Mpc}^2$]}
		\includegraphics[width=0.45\textwidth]{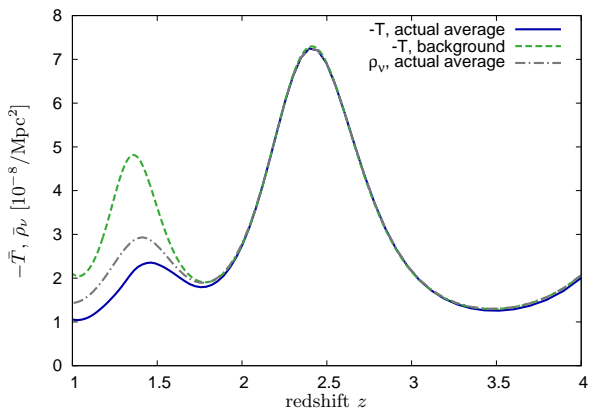}
	\end{center}
	\caption{Inconsistency between the assumed background evolution
	(green dashed) and the actual average taking into account the
	locally varying cosmon field (blue solid). For illustration, the
	figure includes the actual average $\bar \rho_\nu$ of the energy
	density (gray dot--dashed).}
	\label{fig:trace_inconsistency}
\end{figure}

One could ask how large the inconsistency becomes if one neglected the
backreaction effects. For this purpose, we run our simulation without
modifying the background evolution due to the influence of
perturbations.  Technically, we use \eq{trtbackground} for the
background evolution.  In order to quantify the inconsistency, we
measure the average, \eq{trttrue}, in every time step. The comparison
between the two estimates of the same quantity $-\bar T = - {\bar
T}^\alpha_{\ \alpha}$ are shown in \fig{trace_inconsistency}. In
addition, we show the actual average $\bar \rho_\nu$ of the energy
density.

For early times, as long as perturbations are small, the two estimates
agree well. But once non--linear structures form, the suppression of
$-\bar T$ becomes obvious.  This error carries over to the evolution
of $\bar \varphi$ by \eq{bgphi} and thereby to the whole subsequent
evolution of the background.

It is thus important to incorporate the backreaction effects and to
evolve the background and the perturbations in parallel. In every time
step, we thus use the actual average, \eq{trttrue}, for evolving the
cosmon field $\bar \varphi$ by \eq{bgphi}.  

We further see the importance of relativistic corrections.  Without
these, we would have $- {\bar T} = \bar \rho_\nu$.
Figure~\ref{fig:trace_inconsistency} shows that this approximation
becomes invalid at late times.  In order to determine the expansion
rate $\mathcal H$, we measure the actual average $\bar \rho_\nu$ in
our simulation box at every time step, instead of using background
equations. The rate $\mathcal H$ is then obtained from the Friedmann
equation $3 \mathcal H^2 = \sum_{i}^{}\bar \rho_i a^2$.

\subsection{Initial conditions}
\label{sec:initial}

Our simulations use the results from linear perturbation theory to draw initial
conditions. The choice of an
initial redshift $z_i$ is motivated by two restrictions. First, as long as neutrinos
are highly relativistic (for $z \gtrsim 5$), one cannot neglect the higher moments in their
phase--space distribution function. Second, linear perturbation theory becomes
invalid at $z \approx 2$. We choose $z_i = 4$ where the equation of state
$w_\nu$ has fallen to about $10^{-2}$ while the neutrino perturbations are still small.

Although $z_i = 4$ is a good choice for the neutrinos, matter perturbations have
to be treated non--linearly much earlier. We thus start the $N$--body treatment
of matter at $z = 49$. During the stage from $z=49$ to $z=4$, matter evolves
according to gravity on a growing neutrino quintessence background.

Typically, $N$--body codes for CDM simulations
use a displacement field to find growing mode initial conditions for
the particles in the Zel'dovich approximation (see, \eg,
\cite{Efstathiou85, Dolag08}). In the more complicated case of growing
neutrino quintessence, however, we prefer to use the results of the
linear calculation to draw the initial conditions directly. 

\subsubsection{Initial random fields}

Before we distribute neutrino particles, we have to draw random perturbation
fields, namely, the density contrast $\delta_\nu(\eta_i, \vec x)$ and the peculiar velocity field $\vec
v^\text{pec}_\nu(\eta_i, \vec x)$. We focus on
scalar perturbations, \ie, the peculiar velocity $\vec v^\text{pec}_\nu$ is related
to a scalar velocity perturbation $v_s$ via $\vec v^\text{pec}_\nu(\eta, \vec x) =
\vec \nabla v_s(\eta, \vec x)$. 

Let us consider generic scalar perturbations $\delta A(\eta, \vec k)$,
$\delta B(\eta, \vec k)$ in Fourier space. They are solutions of the
perturbation evolution equations as given in \cite{Mota08}. These are
linear and only depend on the absolute value $k=|\vec k|$. Hence, they
allow for basis solutions ${\mathcal A}_k(\eta)$, $ {\mathcal
B}_k(\eta)$.  We consider the adiabatic mode and write
\begin{equation}
	\binom{\delta A(\eta, \vec k)}{\delta B(\eta, \vec k)} =
	\alpha(\vec k)\, \binom{{\mathcal A}_k(\eta)}{{\mathcal
	B}_k(\eta)}. 
	\label{eq:binom}
\end{equation}
The coefficient $\alpha(\vec k)$ characterizes the concrete
realization, whereas the basis solutions ${\mathcal A}_k(\eta)$,
${\mathcal B}_k(\eta)$ are merely attributes of the differential
equations. The two--point statistics is then
\begin{equation}
	\left< \delta A(\eta, \vec k) \delta B^*(\eta, \vec q)\right> =
	{\mathcal A}_k(\eta) {\mathcal B}_{q}(\eta)\left< \alpha(\vec k)
	\alpha^*(\vec q)\right>. 
\end{equation}
Assuming that, at an early cosmological time
$\eta_\text{prim}$, the perturbations were described by the
primordial spectrum $P_\text{prim}(k)$, it follows
\begin{equation}
	\left< \alpha(\vec k)\, \alpha^*(\vec q) \right> = (2 \pi)^3
	P_\text{prim}(k)\, \delta^3(\vec k - \vec q).
	\label{eq:alphastat}
\end{equation}
We assume Gaussian primordial perturbations with a
Harrison--Zel'dovich spectrum
\begin{equation}
	P_\text{prim}(k) = \frac{2 \pi^2}{k^3}\, A_s\, \left(
	\frac{k}{k_\text{pivot}} \right)^{n_s-1},
	\label{eq:prim}
\end{equation}
where $A_s$ denotes the scalar amplitude, $n_s$ the spectral index,
and $k_\text{pivot}$ the pivot scale.

A random realization of the fields $\delta A(\eta_i,\vec x)$ and
$\delta B(\eta_i, \vec x)$ used as initial fields in the simulation is
now obtained in three steps. First, $\alpha(\vec k)$ is drawn as a
Gaussian field with the statistics given in \eq{alphastat}. Second,
the linear code \cite{Mota08} is used to obtain the basis solution
${\mathcal A}_k(\eta_i)$ and ${\mathcal B}_k(\eta_i)$.  Since the
linear code works in synchronous gauge, a gauge transformation to the
Newtonian gauge is necessary (\cf, \eg, \cite{Ma95}). Finally,
\eq{binom} determines the perturbation fields $\delta A(\eta_i, \vec
k)$ and $\delta B(\eta_i, \vec k)$ in Fourier space, which can be
transformed to real space. 

The procedure above also applies to matter starting at $z=49$. A consistent
realization of the two species requires to use the same coefficients $\alpha(\vec
k)$ for both.         

\subsubsection{Discrete fields and particle distribution}

In a numerical implementation, real and Fourier space are discretized.
We will now explain how to draw initial conditions on a discrete grid
$\{\vec k_i\}$.  Therefore, we first have to derive a discretized
version of \eq{alphastat}. In particular, we work with a discrete
Fourier transform (DFT) rather than the continuous version.

First of all, on the right--hand side, the Dirac delta function is
approximated by the Kronecker delta,
\begin{align}
	(2 \pi)^3 P_\text{prim}(k_i)\, \delta^3(\vec k_i - \vec k_j)
	&\approx (2 \pi)^3 P_\text{prim}(k_i)\, \frac{\delta_{ij}}{(\Delta
	k)^3} \nonumber \\ &= P_\text{prim}(k_i)\, L^3\, \delta_{ij},
	\label{eq:kron}
\end{align}
where we have used $\Delta k = 2\pi/L$.  On the left--hand side of
\eq{alphastat}, we replace the quantities $\alpha(\vec k_i)$ by the
discrete Fourier quantities $\tilde \alpha_i$ related to real space by
a DFT.

The relation between $\alpha(\vec k_i)$ and $\tilde \alpha_i$ becomes
obvious when we discretize the Fourier integral. We write $\vec k_i =
(i_1, i_2, i_3)\, \Delta k$ and $\vec x_j = (j_1, j_2, j_3)\, \Delta
x$ with $\Delta x = L / n$ and the number $n$ of cells per dimension,
\ie $n^3 = N_c$. It follows
\begin{align}
	\alpha(\vec k_i) &= \int_{}^{} \dd^3 x \, \alpha(\vec x) \,
	\ee^{-\ii \vec k_i \cdot \vec x_j}
	\nonumber \\
	&\approx \frac{V}{N_c} \sum_{j}^{} \alpha(\vec x_j) \,
	\ee^{-2\pi\ii\,(i_1 j_1 + i_2 j_2 + i_3 j_3) / n}
	\nonumber \\
	&= \frac{V}{N_c}\, \tilde \alpha_i,
	\label{eq:dft}
\end{align}
where we have used the definition of the three--dimensional DFT.

We arrive at the discretized version of \eq{alphastat},
\begin{equation}
	\left< |\tilde \alpha_i|^2 \right> = \frac{N_c^2}{V}
	P_\text{prim}(k_i), 
\end{equation}
expressing the variance of the Gaussian random number $\tilde
\alpha_i$. Only approximately half of the $N_c$ numbers $\tilde
\alpha_i$ are independent. We have to respect the hermiticity
condition ensuring that the transformed field $\alpha(\vec x)$ takes
on only real values. In this way, we obtain our initial random fields
on the grid.    

Once the fields $\delta_\nu(\vec x)$, $\delta_m(\vec x)$, $\vec
v^\text{pec}_\nu(\vec x)$, and $\vec v^\text{pec}_m(\vec x)$ are calculated, we have to
represent them by a distribution of $N$--body particles. We freely
choose a total number of $N$--body neutrino and matter particles,
$N_\nu$ and $N_m$, respectively. 

For the neutrinos, we have to account for thermal velocities $\vec v^\text{th}$, which, at 
$z_i=4$, are small but
not negligible. They are described by a Fermi--Dirac distribution $f_0(\vec
v^\text{th})$, \cf, \eg,
\cite{Ma95}. The total particle velocity is then
$\vec v= \vec v^\text{th} + \vec v^\text{pec}$. We approximate the phase--space
distribution function of neutrinos by
\begin{equation}
	f_\nu(x^i, v_j) = \frac{N_\nu}{V} \, f_0(\vec v - \vec v^\text{pec}_\nu(\vec
	x))\, (1+\delta_\nu(\vec x)).
	\label{eq:fnu}
\end{equation}
In this way, the averaged particle velocities at each cell reproduce the peculiar velocity
field.

The distribution function implies the number density 
\begin{equation}
	n_\nu(\vec x) = \int_{}^{}\dd^3 v\, f_\nu(x^i, v_j) = \frac{N_\nu}{V}
	(1+\delta_\nu(\vec x)). 
	\label{}
\end{equation}
At each grid point $\vec x$, we distribute the rounded number $\lfloor
n(\vec x)\, (\Delta x)^3 \rfloor$ of particles and correct the error
statistically by the addition of further particles. To reduce shot
noise, particles are put at random positions within the pixel volume.
The velocity $\vec v$ of a particle in a pixel $\vec x$ is drawn from the
distribution $f_0(\vec v - \vec v^\text{pec}_\nu(\vec x))$. In order to enforce
the correct local average, we draw particles pairwise with opposite thermal
velocities. 

For matter, the procedure is analogous with the only difference that thermal
velocities are negligible:
\begin{equation}
	f_m(x^i, v_j) = \frac{N_m}{V} \, \delta^3(\vec v - \vec v^\text{pec}_m(\vec
	x))\, (1+\delta_m(\vec x)).
	\label{eq:fm}
\end{equation}

We have checked this procedure by comparing the spectra estimated from
the resulting particle distribution with those obtained by the linear
code. 

\subsection{Numerical issues}
\label{sec:numerical}

The strategies presented in the previous sections take care, in principle, of
all relevant effects in growing neutrino quintessence. As an important caveat,
the method encounters numerical instabilities once very concentrated structures
have formed. We will discuss this problem in the following section. Further, we
will address the influence of the limited resolution on our results.

\subsubsection{Instability for $z\lesssim 1$}
\label{sec:instability}

In \sec{particles}, we collected several
equations that allow us to estimate the potentials $\Psi$, $\Phi$, and
$\delta \varphi$ from the distribution of particles. These
calculations, however, already require knowledge of the potentials,
\eg via the metric coefficients, \cf \eq{detgtilde}. We cannot resolve these mutual dependencies exactly. 

As long as the time steps are chosen small enough, it is a good
approximation to use, in these cases, the potentials of the previous
step. This is the procedure we generally apply. 

This can fail, however, if small errors accumulate in such a way that
the evolution becomes significantly inaccurate or even unphysical. In
fact, once the formation of neutrino structures is advanced, this
problem occurs in the estimation of $\delta \varphi$. This limits our
capability to simulate growing neutrino quintessence for late
cosmological times, $z \lesssim 1$. 

In more detail, \eq{pertphi} for $\delta \varphi$ contains a source
term $\delta T = \delta {T^\alpha}_\alpha$, which, in turn, is
sensitive to the local neutrino mass variations $\propto \exp(-\beta
\delta \varphi)$. Formally, \eq{pertphi} can be written in the
abstract form
\begin{equation}
	\Delta \delta \varphi(x) = f(\delta \varphi(x); x) ,
	\label{eq:pde}
\end{equation}
with a highly non--linear function $f$. 

Attempts to discretize the Laplacian and to apply Newton's method to
directly solve this non--linear equation are numerically intractable
due to the large number of cells.

Another approach is to consider the iteration
\begin{equation}
	\Delta \delta \varphi^{(n+1)}(x) = f(\delta \varphi^{(n)}(x); x).
	\label{eq:iteration}
\end{equation}
If it converges to a fixed point $\delta \varphi^*$, this is clearly a
solution of \eq{pde}. As a starting point $\delta \varphi^{(0)}$, one
may use the result of the previous time step. 

The convergence of the sequence $\delta \varphi^{(n)}$, however, is
not guaranteed. We apply this iterative scheme until $z=1$, since, for
later times, the sequence starts to diverge.

In order to get a rough idea how the behavior of the sequence depends
on the physical conditions, we consider a very simplified
configuration.   We assume the idealized situation in which all the
neutrinos in the volume $V$ are concentrated in a dense structure of
comoving size $l$ small enough to be approximated by a Dirac delta
shape. Let us fix the origin of the coordinates $\vec x$ at the
position of the structure.  The structure is assigned a total mass
$M^{(n)} \propto \exp(-\beta \delta \varphi^{(n)}(0))$ in the $n$--th
iteration. For convenience, we take care only of the leading effects,
\ie, we drop the metric perturbations and neglect
$V_{,\varphi\varphi}$ in \eq{pertphi}.

The resulting expression for $T$ reads
\begin{equation}
	T = \frac{1}{a^3} \,
	\frac{M^{(n)}}{\gamma_\text{eff}}\, \delta^3 (\vec x),
	\label{}
\end{equation}
where $\gamma_\text{eff}$ arises from the Lorentz factors of the
neutrinos inside the structure. The iteration prescription,
\eq{iteration}, in Fourier space yields, for $\vec k \not= 0$,
\begin{equation}
	\delta\varphi^{(n+1)}_{\vec k} = \frac{1}{k^2}\,\beta
	a^2\,\frac{M^{(n)}}{\gamma_\text{eff}}.
	\label{}
\end{equation}
Transformed back to real space and evaluated at the structure position
$\vec x = 0$,
\begin{equation}
	\delta\varphi^{(n+1)}(0) = \frac{\beta}{2\pi a} \,
	\frac{M^{(n)}}{l\,\gamma_\text{eff}},
	\label{}
\end{equation}
where we have performed a cut--off of the Fourier integral at the
scale of the structure size $k_\text{max} = \pi / l$. This value
determines the next mass estimate $M^{(n+1)} \propto \exp(-\beta
\delta \varphi^{(n+1)}(0))$.

If we had started with the correct solution $\delta \varphi^*$, the
mass $M^*$ would remain fixed in the iteration. Let us assume we had
started with a small error, $M^{(n)} = M^* \, (1 +
\varepsilon^{(n)})$. Then we find, at linear order in
$\varepsilon^{(n)}$, the size of the error in the next step,
\begin{equation}
	\varepsilon^{(n+1)} = - \frac{\beta^2}{2\pi a} \,
	\frac{M^*}{l\, \gamma_\text{eff}}\, \varepsilon^{(n)}.
	\label{}
\end{equation}
The iteration can only converge if the errors decrease, \ie, if the
absolute value of the prefactor is smaller than unity.

Of course, the above discussion is only illustrative. Nevertheless, it
clarifies that the strong coupling $\beta$ and large concentrations
$M^*/l$ of neutrinos in structures are working against the iteration.

\subsubsection{Resolution}

The equations of motion for the effective neutrino particles, \eq{eomnu},
are not reproduced by mere two--body forces. Instead, they require the knowledge
of the fields $\Psi(\vec x)$, $\Phi(\vec x)$, and $\delta \varphi(\vec x)$ on
a grid. This differs significantly from pure CDM simulations, which obtain a
high resolution by using two--body interactions for small--scale forces as in
{\sc gadget}--{\small 2} \cite{Springel05}. Since we do not employ an adaptive mesh, our
resolution is limited by the constant size of a grid cell $\Delta x$.  While
this clearly affects the accuracy of our method at small scales (comparable to
$\Delta x$), our approach still seems appropriate for studying the model 
on large scales at which neutrino structures first form. 

\begin{figure}[htb!]
	\begin{center}
		\psfrag{xlabel}[B][c][.8][0]{scale $k$ [$h/$Mpc]}
		\psfrag{ylabel}[T][c][.8][0]{neutrino spectrum $\Delta_\nu$}
		\includegraphics[width=0.45 \textwidth]{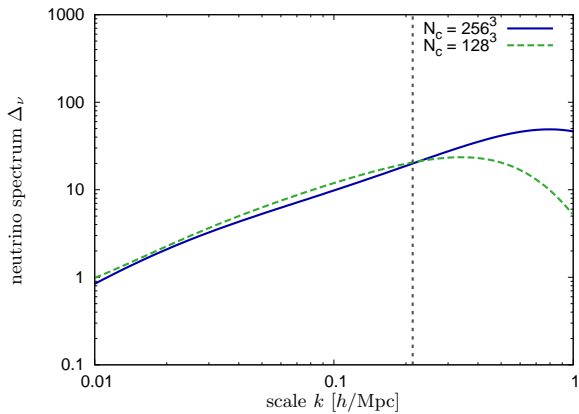}
	\end{center}
	\caption{The neutrino spectrum $\Delta_\nu$ at $z=1$ for two
	different resolutions. The vertical dashed line marks the scale $1/\Delta x \approx
	0.25\, h/$Mpc roughly corresponding to the cell size of the low--resolution
	grid.}
	\label{fig:resolution}
\end{figure}

As a check, we have run our simulation with a lower resolution, $N_c =
128^3$, corresponding to twice a bigger cell size $\Delta x \approx 4\,
h^{-1}$Mpc. Apart from $N_c$, all other
parameters as well as the initial random realizations have been chosen
identical.
In \fig{resolution}, we show the dimensionless neutrino spectrum,
\begin{equation}
  \Delta_\nu^2 (k) = \frac{k^3}{2 \pi^2}\, P_\nu(k), 
  \label{eq:Delta_nu}
\end{equation}
at redshift $z=1$ obtained from the two simulation runs.
As expected, the reduction of $N_c$ leads to a loss of power on small scales. On
the other hand, we find satisfying agreement on large scales $k \lesssim
1/\Delta x \approx 0.25\, h/$Mpc. These results indicate that we can obtain a robust picture of
large--scale neutrino clustering from our simulations. When studying the
properties of individual neutrino structures in \sec{individual}, the resolution
of small scales will be more important. We will then again use the
low--resolution simulation
with $N_c=128^3$ to quantify the effect.

\section{Neutrino structures}
\label{sec:neutrino}

Using the strategies presented in \sec{strategy}, we are able to study
growing neutrino quintessence at the non--linear level. We first
investigate qualitatively how the strong fifth force leads to the
formation of large neutrino structures, \sec{growth}. Within these
structures, the local mass variation becomes an important effect as we
will see in \sec{individual}. The effects of neutrinos accelerated to
relativistic velocities are presented in \sec{relativistic}.  Since
the cosmological neutrinos are not directly observable, the indirect
influence of neutrino structures on cosmological observables is
essential and will be the topic of \sec{impact}.

The model parameters used in our simulation are listed in \tab{model}.
\begin{table}[htb!]
	\begin{center}
		\begin{tabular}{ccc}
			Model & & Cosmology \\
			\hline
			$\beta =-52$ & & $\mathcal H_0 = 70$~km/s~Mpc$^{-1}$ \\
			$\alpha = 10$ & & $A_s = 2.3 \times 10^{-9}$\\
			$m_\nu^{0} = 2.3$~eV & & $k_\text{pivot} =
			0.05$~Mpc$^{-1}$\\
			$\Omega_\nu^0 = 0.15$ & & $n_s = 0.96$ \\
			$\Omega_\varphi^0 = 0.60$ & &  \\
		\end{tabular}
		\caption{Parameters for growing neutrino quintessence and
		primordial perturbations.}
		\label{tab:model}
	\end{center}
\end{table}
These are exemplary parameters resulting in a rather large
present--day neutrino mass. Very different parameter choices are
possible, including a time--varying coupling constant $\beta =
\beta(\varphi)$. The values $m_\nu^0$, $\Omega_\nu^0$, and
$\Omega_\varphi^0$ would be reached if no backreaction effects were
taken into account. They do not characterize the state at $z = 0$ of
the full cosmological evolution including backreaction.

\subsection{Growth of structure}
\label{sec:growth}

\begin{figure*}[htb!]
	\begin{center}
		\includegraphics[height=0.8 \textheight]{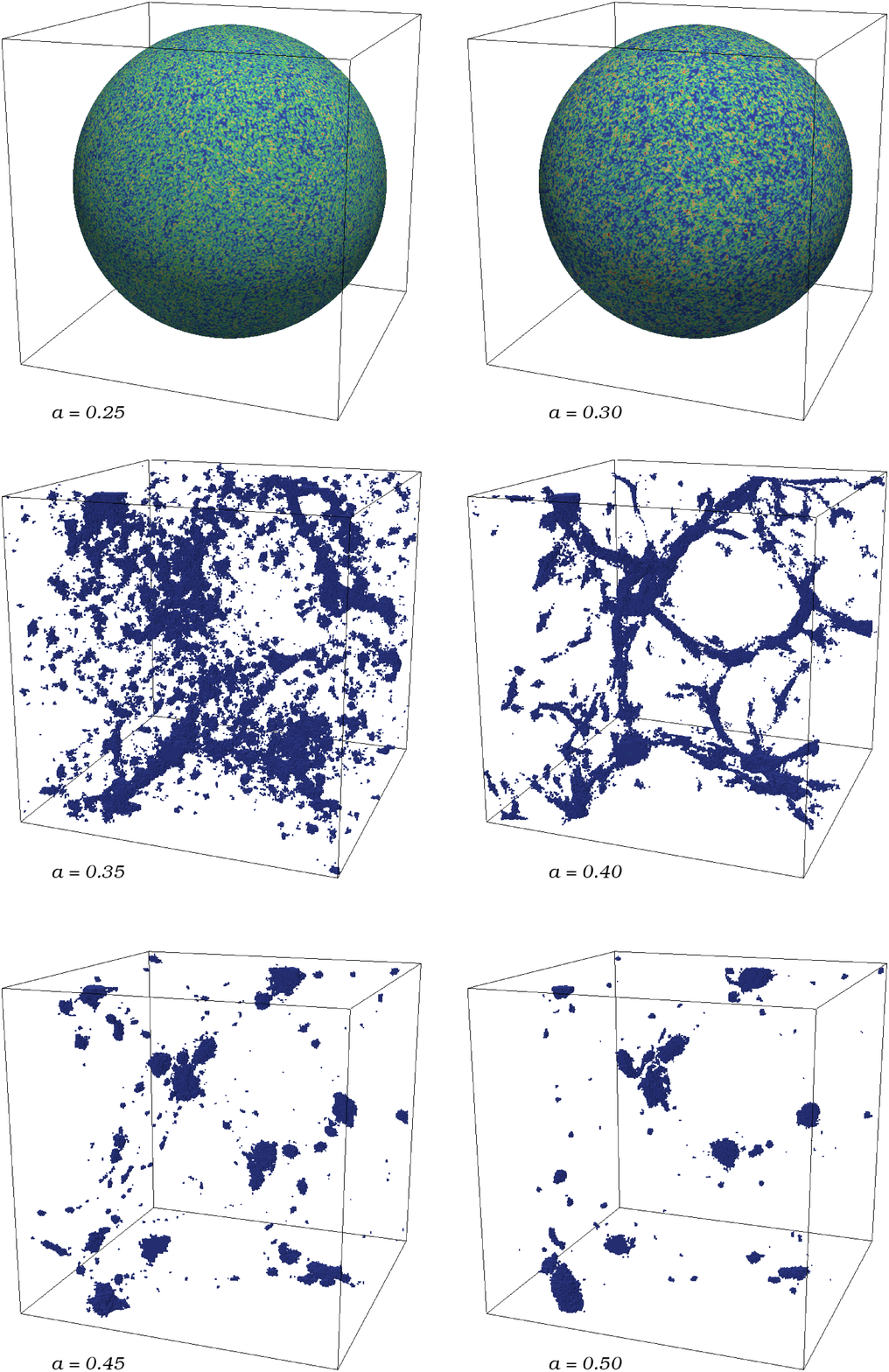}
	\end{center}
	\caption{The evolution of non--linear neutrino structures in the
	simulation box, $L = 600\,h^{-1}$Mpc. The lower four figures show
	regions in the simulation box where the number density of
	neutrinos is a factor of $\geq 5$ higher than in the background.
	We see the process of continuing concentration. In the beginning,
	the structures were still linear. We show, in the upper figures, a
	two--dimensional section of the number density (the color
	range---blue to red---goes from $0$ to $5$ times the background
	value).}
	\label{fig:film}
\end{figure*}

We run the simulation with the cosmological parameters listed in
\tab{model} and the simulation parameters of \tab{values}.  Recalling
that the cosmon--mediated fifth force felt by the neutrinos is a
factor of about $2\beta^2 \approx 5\times 10^3$ stronger than gravity,
it is expected that large non--linear neutrino structures will form.
The numerical results we get in our simulation agree with this
expectation. We can follow the precise time evolution of the structure
formation process. We give an overview of the evolution from $a =
0.25$ to $a = 0.5$ in \fig{film}.
The images show snapshots of the number density field $n_\nu(\vec x)$ of
neutrinos in the simulation box (periodic boundary conditions) at
intervals $\Delta a = 0.05$. We will now discuss the main stages of
neutrino structure formation.

Until $a \approx 0.3$, no large non--linear structures have formed.
The perturbations can still be described linearly, and our results
reproduce the linear calculation \cite{Mota08}. We give an impression
of these perturbations and their growth by showing a spherical,
two--dimensional section of the three--dimensional field.  Although
mainly linear, the growth already is very fast. From $a = 0.25$ to $a
= 0.30$, the perturbations have grown by a factor of about $3$ to $4$.
The particle velocities, however, are still far from relativistic. The
Newtonian limit would be applicable, and we will also see that
backreaction effects are small.

From $a \approx 0.3$ to $0.4$, large--scale non--linear
inhomogeneities are forming. Most cells in the simulation box are
already empty of effective neutrino particles, which increasingly
concentrate in large and dense filament--like regions. During this
process, the velocities of effective neutrino particles increase and
often reach relativistic values. The Newtonian limit breaks down, and
the fully relativistic equation of motion becomes essential. The large
density inhomogeneities lead to non--negligible local variations
$\delta\varphi$ of the cosmon and in turn to important local mass
variations. These modified masses together with the Lorentz factors
$\gamma > 1$ cause backreaction effects according to
\sec{backreaction}. In fact, we shall find in
\sec{backreactionandquintessence} that $a \approx 0.4$, or $z \approx
1.5$ is a characteristic time where the evolution of the cosmological
background has started to feel the impact of large inhomogeneities.

After $a \approx 0.4$, approximately spherical neutrino structures
form (as already predicted in \cite{Brouzakis07}), mainly at the
intersections of the large filament--like non--linearities. In the
course of the evolution, these structures become even increasingly
concentrated and spherical. The spherical shapes can be regarded as a
hint for virialization. In the simulation cells, the number
densities of effective particles exceed the average value by factors
up to $10^5$. The structures are, however, very different from matter
structures. They produce local cosmon inhomogeneities $\delta \varphi$
resulting in locally varying neutrino masses. In the structures'
centers, the neutrino mass is heavily suppressed, \cf
\sec{individual}.

In \sec{numerical}, we explained that our strategy to calculate the
cosmon perturbations $\delta \varphi$ becomes problematic once the
concentration of neutrino structures becomes very large. For this
reason, we cannot follow the subsequent cosmological evolution for $a
> 0.5$, \ie $z < 1$. We expect the overall picture to remain stable,
\ie a collection of neutrino structures with very large
central overdensities. Due to the attractive fifth force, some of
these structures are expected to merge if they are close enough to
each other.

\subsection{Individual structures}
\label{sec:individual}

\begin{figure}[htb!] 
	\begin{center}
		\includegraphics[width=0.40\textwidth]{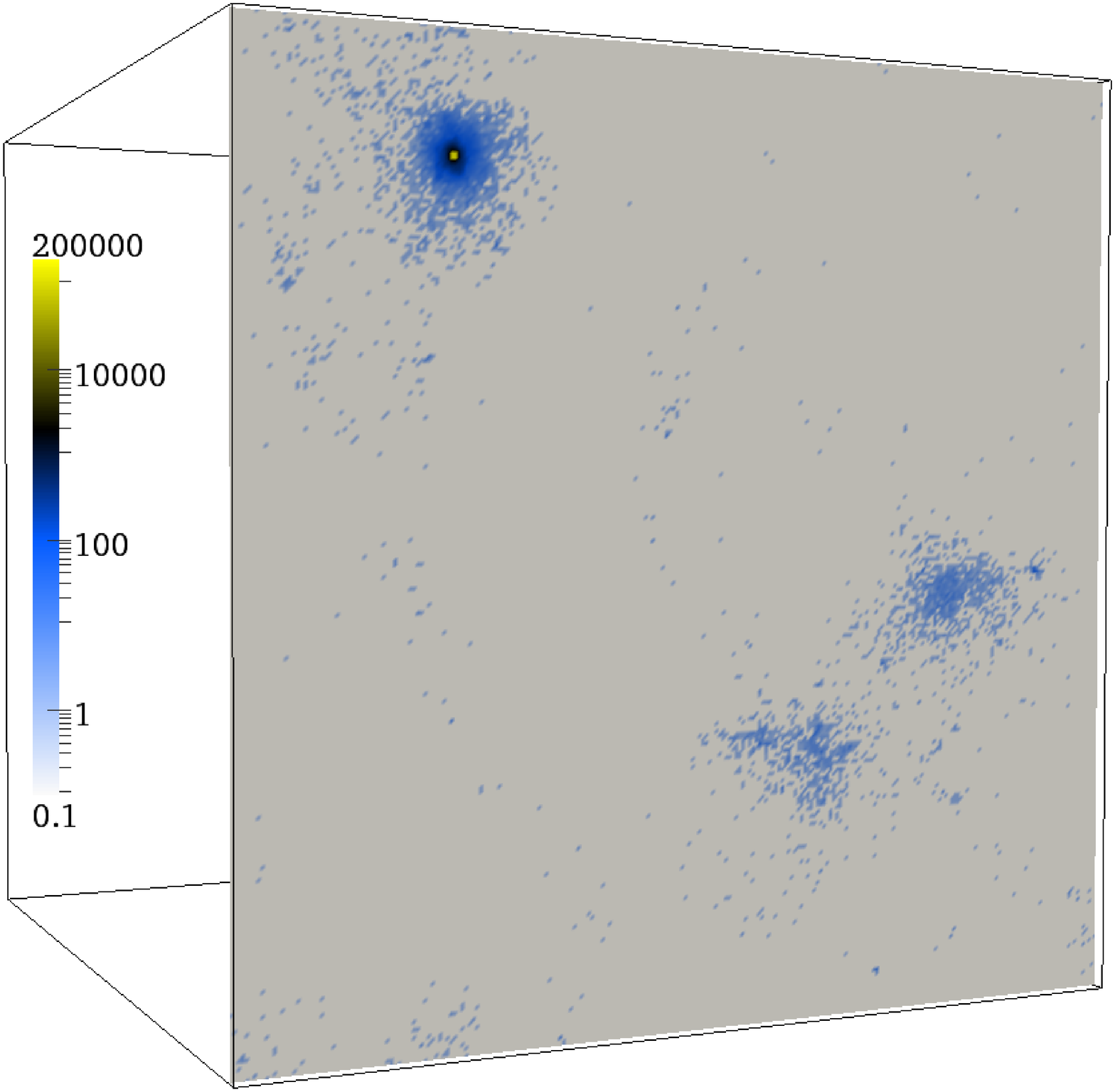}
		\psfrag{xlabel}[B][c][.8][0]{physical radial distance $r_\text{ph}$ [$h^{-1}$Mpc]}
		\psfrag{ylabel}[T][c][.8][0]{number density profile $
		n_\nu/{\bar n_\nu}$}
		\includegraphics[width=0.45\textwidth]{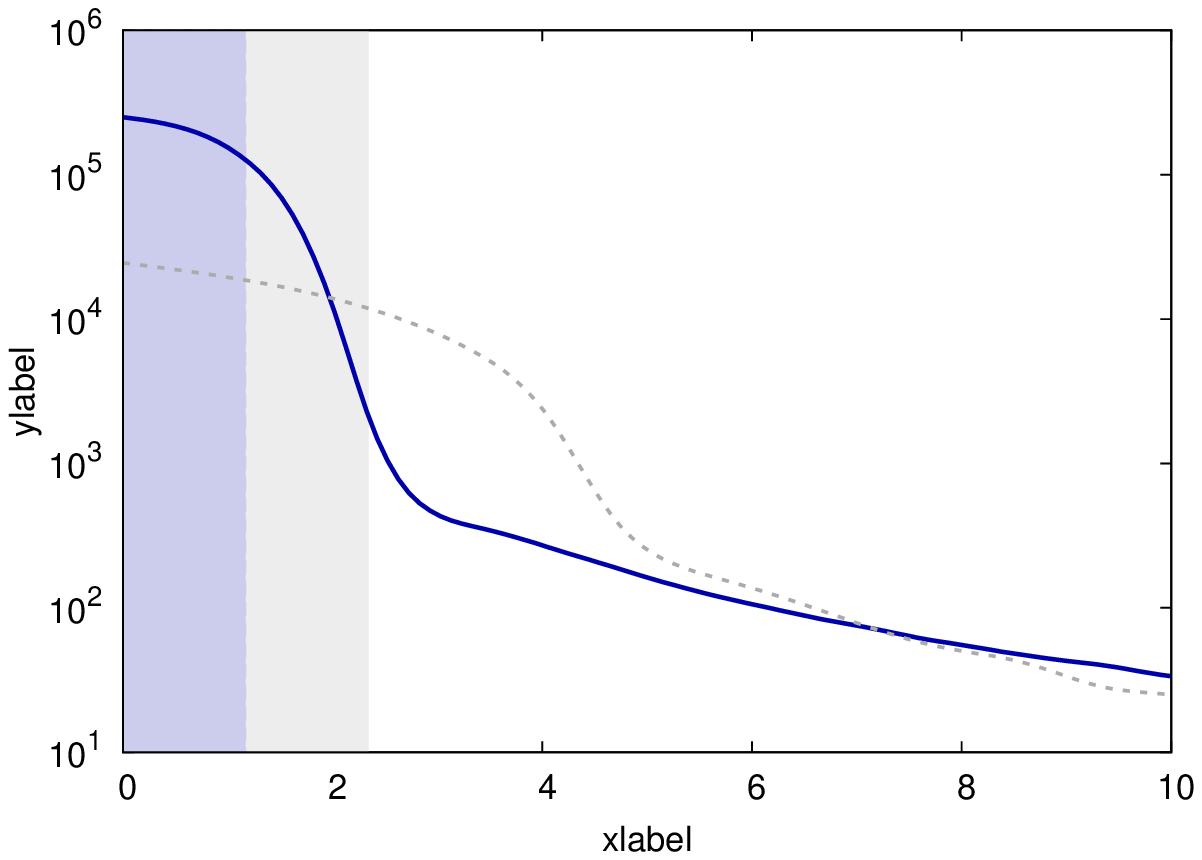}

		\psfrag{xlabel}[B][c][.8][0]{physical radial distance $r_\text{ph}$ [$h^{-1}$Mpc]}
		\psfrag{ylabel}[B][c][.8][0]{mass profile ${
		m}_\nu$ [eV]}
		\includegraphics[width=0.45\textwidth]{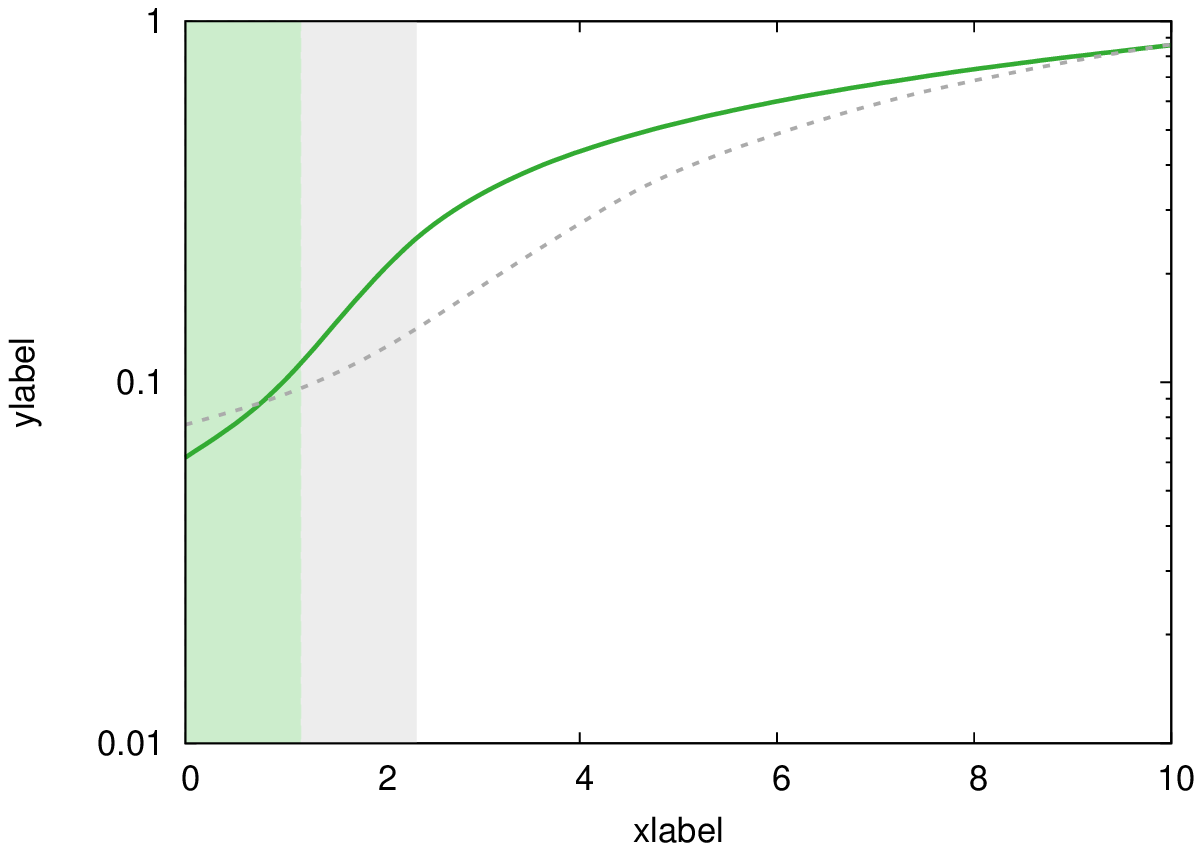}
	\end{center}
	\caption{The uppermost figure shows a slice through the number density
	field of neutrinos (in multiples of the average $\bar n_\nu$). The two lower figures show
	the number density profile $n_\nu(r)$ and the mass profile $m_\nu(r)$ of
	the structure. The dashed lines show the results of a simulation with a lower resolution.
	The shaded regions indicate the size of
	a grid cell for each resolution.}
	\label{fig:slice}
\end{figure}

We have found several distinct neutrino structures in our simulation
box at $z=1$.  In the following, we have a closer look at one of the
most pronounced structures. In \fig{slice}, we show a two--dimensional
slice through our simulation volume located at the center of the
structure. 
Since the structure is almost spherically symmetric, we can
characterize its shape by a radial profile.  We show the neutrino
number density and the local neutrino mass as a function of the
physical radial distance $r_\text{ph}$ from the center of the structure.  A 
concentrated core is clearly visible with a neutrino number density contrast
of $2 \times 10^5$ in the central cell. 

A very interesting phenomenon of growing neutrino quintessence are the
local neutrino mass variations. Since the cosmon--mediated  fifth
forth leads to a substantial neutrino clustering in regions of
negative cosmon perturbations $\delta \varphi$, the neutrino mass is
typically strongly suppressed in overdense regions. This suppression
is clearly visible inside our exemplary structure, \cf \fig{slice}.
We find that the neutrino mass at the innermost core of the structure
is approximately one order of magnitude smaller than outside the
structure. This result stresses the relevance of the local cosmon
perturbations. It remains an open question how the structures evolve for $z < 1$.
The idea that the neutrino mass profile inside the structures may
become constant and independent from the evolution of the cosmological
background has been raised in \cite{Nunes11}. 

Our simulation method does not resolve the dynamics below the size of a cell,
which is around $2\, h^{-1}$Mpc in comoving units.
Hence, at small distances from the center, the quantitative results are affected
by rather large errors. In order to estimate the uncertainties, we show the
corresponding results obtained from a simulation with a twice as large cell size
(dashed lines). The number density profile shows a similar shape but is significantly less concentrated. Still, the mass suppression reaches similar values.    

\begin{figure}[htb!]
	\begin{center}
		\psfrag{xlabel}[B][c][.8][0]{physical radial distance $r_\text{ph}$ [$h^{-1}$Mpc]}
	\psfrag{ylabel}[T][c][.8][0]{number density profile $n_\nu/\bar
	n_\nu$}
	\includegraphics[width=0.45\textwidth]{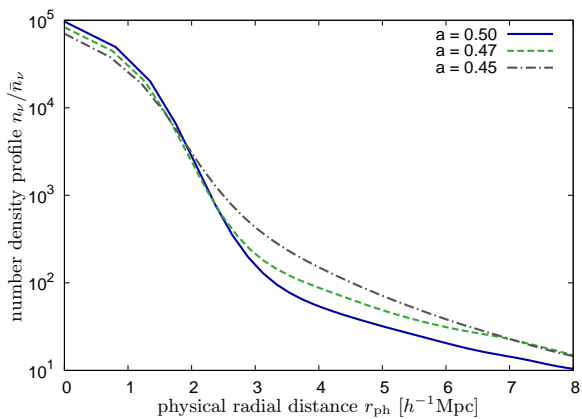}
	\end{center}
	\caption{The figure shows the number
	density profile $n_\nu$ of an isolated neutrino structure at
	$a=0.50$ (blue solid), $a=0.47$ (green dashed), and $a = 0.45$
	(gray dot--dashed) as a function of the
	physical distance from its center.}
	\label{fig:virial} 
\end{figure}

We have seen that the neutrino structures are still becoming
increasingly concentrated in the final stage of our evolution (\cf
\fig{film}). This is most visible for the largest structures and less
significant for smaller ones. As an example, we show how the number
density profile of one of the smaller, but still very pronounced
structures evolves from $a=0.45$ to $a=0.5$ in \fig{virial}. We have
verified that the structure is not undergoing merging processes in the
considered time range.
Apart from a moderate transfer of neutrinos from the outer regions of
the structure to its inner core, no significant changes in the profile
are visible.

\subsection{Relativistic effects}
\label{sec:relativistic}

The motion of the effective neutrino particles in our simulation
respects the relativistic laws of motion, \cf \sec{neutrinodynamics},
which remain valid for velocities approaching the speed of light. This
is essential since the neutrinos feel a strong acceleration due to the
fifth force. We observe in our simulations that the Newtonian laws of
motion assuming small velocities become inadequate as soon as
non--linear clustering begins, $z \approx 2$ to $1.5$, \cf
\sec{growth}.

The importance of the relativistic treatment becomes manifest in the
evolution of the average equation of state $w_\nu = \bar p_\nu / \bar
\rho_\nu$. With the help of the methods explained in \sec{particles},
we can calculate the average pressure $\bar p_\nu$ and energy density
$\bar \rho_\nu$ of the neutrino fluid from the distribution of
particles in our simulation. Performing the average over the
contributions of single particles $p$ to the density $\bar \rho_\nu = -
{\bar T}^0_{\ 0}$, \eq{t00nu}, yields
\begin{equation}
	\bar \rho_\nu
	= -\frac{\int_{V}^{} \dd^3 x\, \sqrt{\tilde g}\,
	{T^0}_0}{\int_{V}^{} \dd^3 x\, \sqrt{\tilde g}}
	= \frac{1}{a^3 V} \sum_{p}^{}
	\gamma_p M_\nu(\varphi(\vec \xi_p)).
	\label{}
\end{equation}
The average pressure $\bar p_\nu$ then follows from the calculation of
${\bar T} = {\bar T}^\alpha_{\ \alpha}$ by \eq{trttrue}, $\bar p_\nu =
(\bar \rho_\nu +{\bar T})/3$. 

\begin{figure}[htb!]
	\begin{center}
		\psfrag{xlabel}[B][c][.8][0]{redshift $z$}
		\psfrag{ylabel}[B][c][.8][0]{average neutrino equation of
		state $w_\nu$}
		\includegraphics[width=0.45\textwidth]{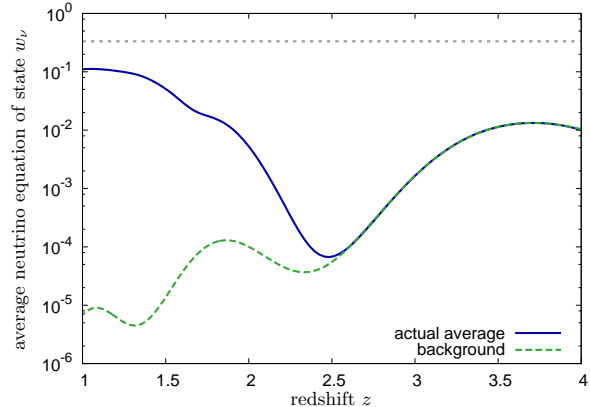}
	\end{center}
	\caption{The increase of the equation of state $w_\nu$ due to
	relativistic velocities in the structure formation process. The
	dashed horizontal line marks the limit $w = 1/3$ for highly relativistic
	particles.}
	\label{fig:wnu}
\end{figure}

The evolution of the equation of state $w_\nu$ is shown in \fig{wnu}. 
The figure shows a steep increase of $w_\nu$, growing several orders
of magnitude between $z=2.5$ and $z=1$, exceeding the value of $w_\nu \approx 0.1$.  
The velocities of many neutrino particles in the box are close to the speed of
light at this stage of the evolution.  The figure also shows the evolution of
$w_\nu$ obtained by solving the homogeneous background equations, \ie neglecting
backreaction effects. If neutrinos were not accelerated by the cosmon--mediated
fifth force, their equation of state would continue to decrease. The pronounced
oscillatory features are attributed to the neutrino mass variation; the cosmon
field oscillates around the minimum of its effective potential (\cf
\cite{Wetterich07}).  

We observe a striking discrepancy between the background calculation and the
evolution of the actual average. The consequences for the evolution of the
background cosmon field $\bar \varphi$ will be discussed in
\sec{backreactionandquintessence}. 

\begin{figure}[htb!]
	\begin{center}
		\includegraphics[width=0.40\textwidth]{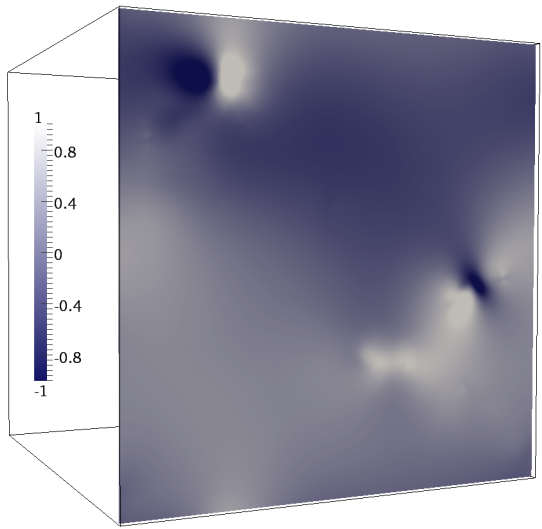}
	\end{center}
	\caption{The difference $\Phi - \Psi$ of the two gravitational
	potentials, scaled by a factor of $10^5$, at $z = 1$.} 
	\label{fig:phi_minus_psi}
\end{figure}

The presence of relativistic neutrinos is also interesting with
regard to gravity. Our tools allow us to calculate the anisotropic
stress induced by neutrinos, which is the source for the difference
between the two gravitational potentials $\Psi$ and $\Phi$,
\sec{particles}. We find that the effect is most pronounced in the
vicinity of neutrino structures and becomes negligible on very large
scales. In \fig{phi_minus_psi}, we show the field $\Phi - \Psi$ at $z=1$ on
the same two--dimensional slice already used in \fig{slice}. 
In the regions of neutrino overdensities, the field $\Phi - \Psi$ 
shows anisotropic patterns with amplitudes of the order $10^{-5}$. This is,
compared to $\Phi$ and $\Psi$ themselves, a $1\%$ to $10\%$ effect. 

\section{Impact on dark energy and matter}
\label{sec:impact}

Once the dynamics of growing neutrino quintessence can be described
reliably, a confrontation with observational data is the next step.
At the current stage, our simulation method allows us to follow the
cosmological evolution only until $z \approx 1$, and we have not explored
the parameter space of growing neutrino quintessence. It is thus not
yet possible to provide constraints on the model parameters.

Nonetheless, it is insightful to discuss some remarkable effects
linked to observations. This regards the expansion history of the
Universe (depending on the perturbation evolution due to strong
backreaction), \sec{backreactionandquintessence}, and the evolution of
matter perturbations, \sec{matter}.

\subsection{Backreaction and quintessence}
\label{sec:backreactionandquintessence}

We argued in \sec{backreaction} that two essential effects in the
structure formation process lead to an important backreaction of the
perturbations on the background evolution. These are the systematic
suppression of neutrino masses within structures, $m_\nu(\varphi(\vec
x)) < m_\nu(\bar \varphi)$, and the relativistic Lorentz factors,
$\gamma > 1$. Numerically, we have seen that these effects are indeed
strong. The neutrino mass inside structures is suppressed by an order
of magnitude, \fig{slice}, and the neutrino equation of state $w_\nu$
grows considerably at late times, \fig{wnu}.
Consequently, the true averaged energy--momentum tensor ${\bar
T}^\alpha_{\ \beta}$ of the neutrino fluid significantly differs from
the idealized calculation based on a homogeneous fluid of
non--relativistic neutrinos.

The strong backreaction is not only interesting in its own right.  In
growing neutrino quintessence, the evolution of the dark energy is
intimately connected, and hence sensitive, to the evolution of the
neutrinos, \cf \sec{growing}. In particular, the cosmological event of
the neutrinos becoming non--relativistic serves as a trigger stopping
the further evolution of the cosmon and leading to an epoch of
accelerated expansion.  We have seen, however, that the fifth force
accelerates neutrinos again to relativistic velocities in the course
of the non--linear evolution. This reduces the neutrinos' capability
of stopping the cosmon evolution. The onset of accelerated expansion
is thus expected to shift to later times.

More precisely, the source term in the background evolution of the
cosmon in \eq{bgphi} is proportional to the trace ${\bar T}$. This
trace was shown to be a sum over particle contributions $\propto m_\nu
/ \gamma$, \eq{trttrue}. So, both the effect of relativistic
particles, $\gamma > 1$, and the mass suppression go into the same
direction.

\begin{figure}[htb!]
	\begin{center}
		\psfrag{xlabel}[B][c][.8][0]{redshift $z$}
		\psfrag{ylabel}[B][c][.8][0]{deceleration parameter $q$}
		\includegraphics[width=0.45\textwidth]{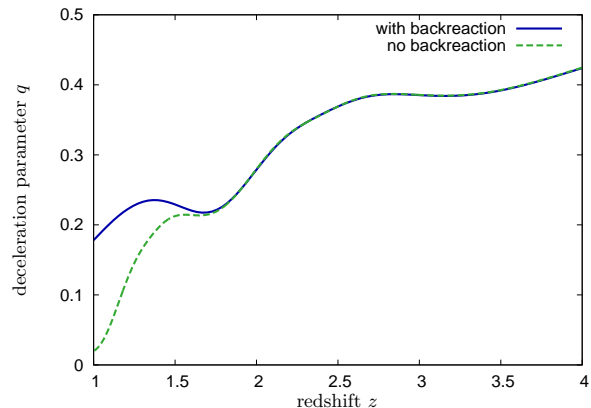}
	\end{center}
	\caption{Evolution of the deceleration parameter $q$ in our
	simulation---including the non--linear backreaction
	effects---(blue solid), compared to the result obtained by the
	background equations (green dashed).}
	\label{fig:deceleration}
\end{figure}

For a quantitative investigation, we calculate the deceleration
parameter $q = - a''\, a / {a'}^2 + 1$. For the expansion of the
Universe to accelerate, $q$ has to take negative values. We show the
evolution of $q$ measured in our simulation, \ie, including
backreaction according to \eq{trttrue}, compared with the expectation
based on the a--priori averaged background equations, \cf
\eq{trtbackground}, in \fig{deceleration}.
We observe that the deceleration parameter, already very close to zero
without backreaction, is indeed still at a relatively large value $q
\approx 0.2$ at $z = 1$. Although our simulation ends at $z = 1$, we
expect the phase of accelerated expansion to start later
due to backreaction effects. This is because we can expect that the
two main contributions, the mass suppression and the relativistic
Lorentz factors, remain important in the subsequent evolution.

By means of Friedmann's equations, we can express the deceleration
parameter $q$ as a sum over contributions of the different species $i$,
\begin{equation}
	q = \frac{a^2}{6 \mathcal H^2}\,\sum_{i}^{} (\bar
	\rho_i + 3 \bar p_i).
	\label{}
\end{equation}
The cosmon contribution to the deceleration thus is
\begin{equation}
	\frac{a^2}{6 \mathcal H^2} (\bar \rho_\varphi + 3 \bar p_\varphi)
	\propto \bar \varphi'^2 - a^2 V(\bar \varphi),
	\label{eq:qphi}
\end{equation}
which is a comparison of the kinetic energy and the potential energy
(the potential is normalized to $V(0) = 0.9\times 10^{-120} M_P^4$
with the reduced Planck mass $M_P$). We can use $\overline{V(\varphi)}
= V(\bar \varphi)$ in linear approximation in $\delta \varphi$. The
difference between the actual average $\overline{V(\varphi)}$ and
$V(\bar \varphi)$ is way below the percent level.

\begin{figure}[htb!]
	\begin{center}
		\psfrag{xlabel}[B][c][.8][0]{redshift $z$}
		\psfrag{ylabel}[B][c][.8][0]{cosmon equation of state
		$w_\varphi$}
		\includegraphics[width=0.45\textwidth]{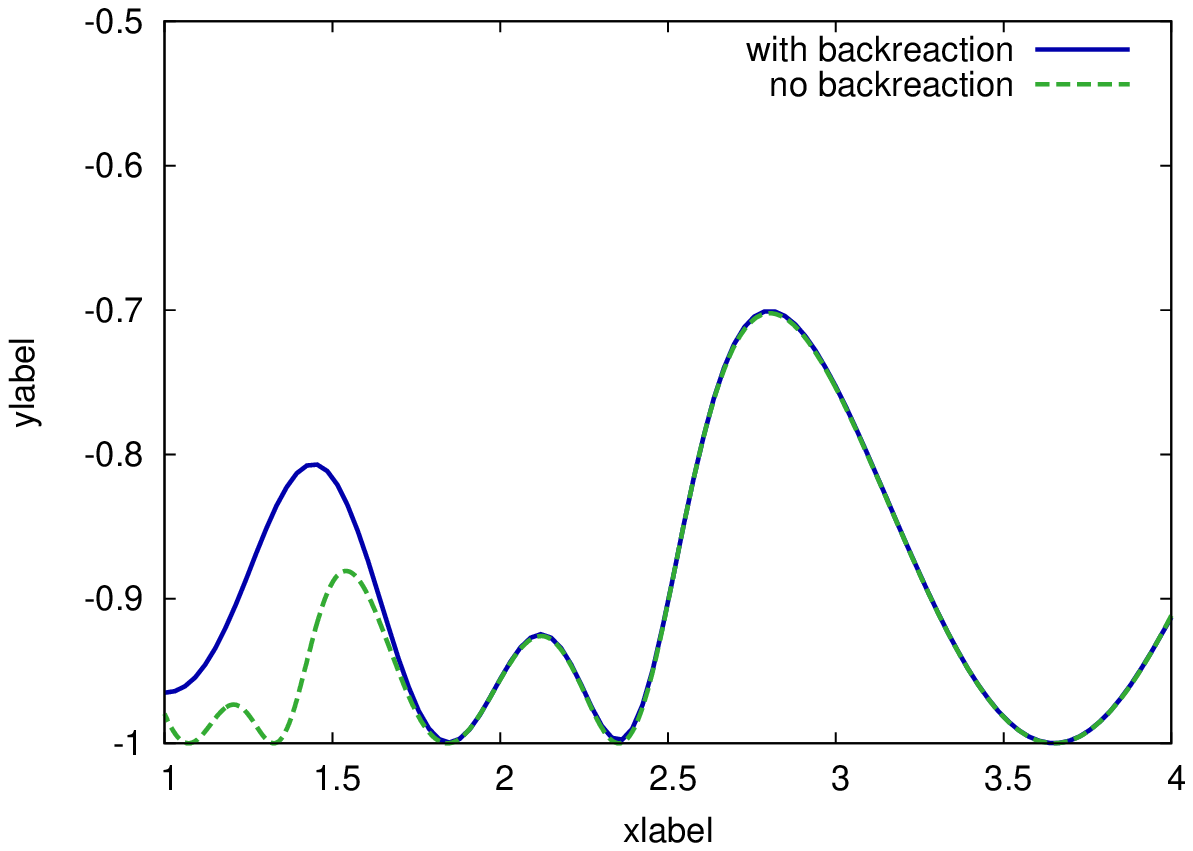}
		\psfrag{xlabel}[B][c][.8][0]{redshift $z$}
		\psfrag{ylabel}[B][c][.8][0]{background cosmon $\bar \varphi$}
		\includegraphics[width=0.45\textwidth]{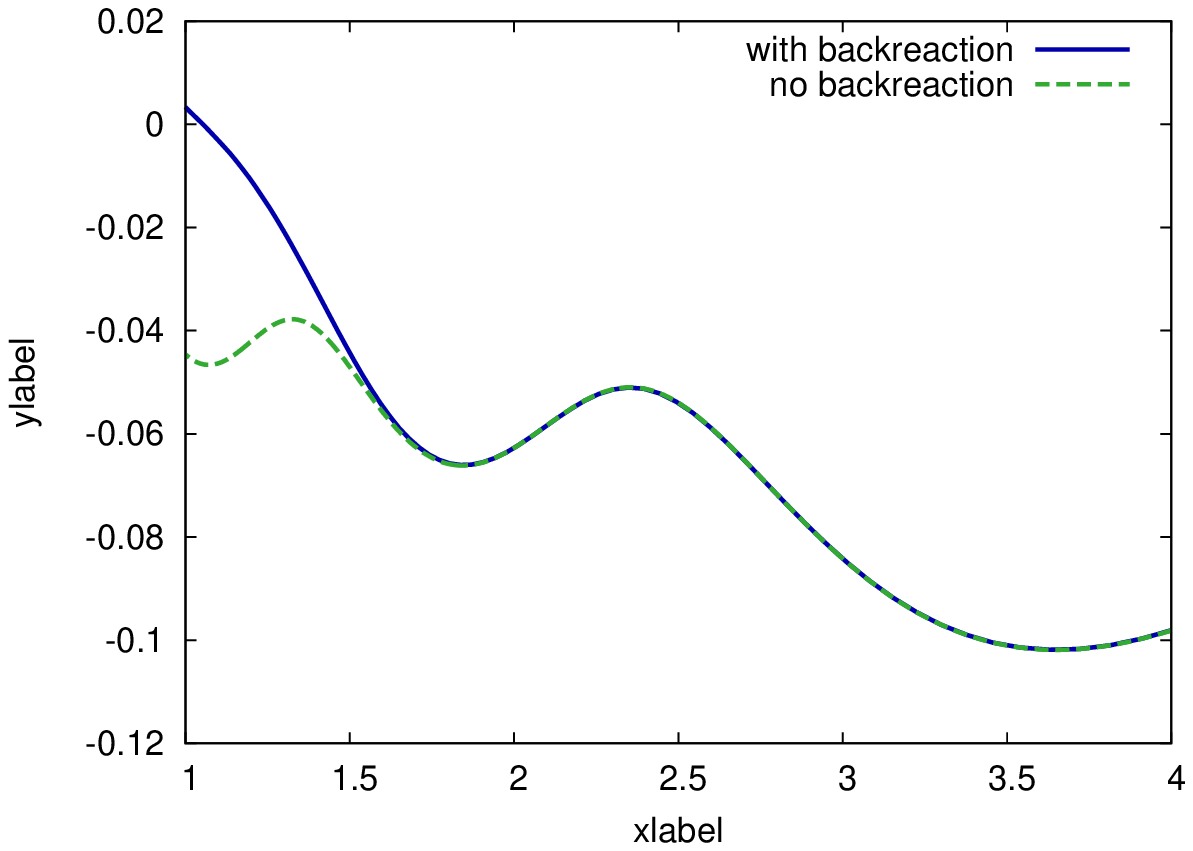}
		\psfrag{xlabel}[B][c][.8][0]{redshift $z$}
		\psfrag{ylabel}[B][c][.8][0]{time derivative $\bar \varphi'$
		[$10^{-5} \text{Mpc}^{-1}$]}
		\includegraphics[width=0.45\textwidth]{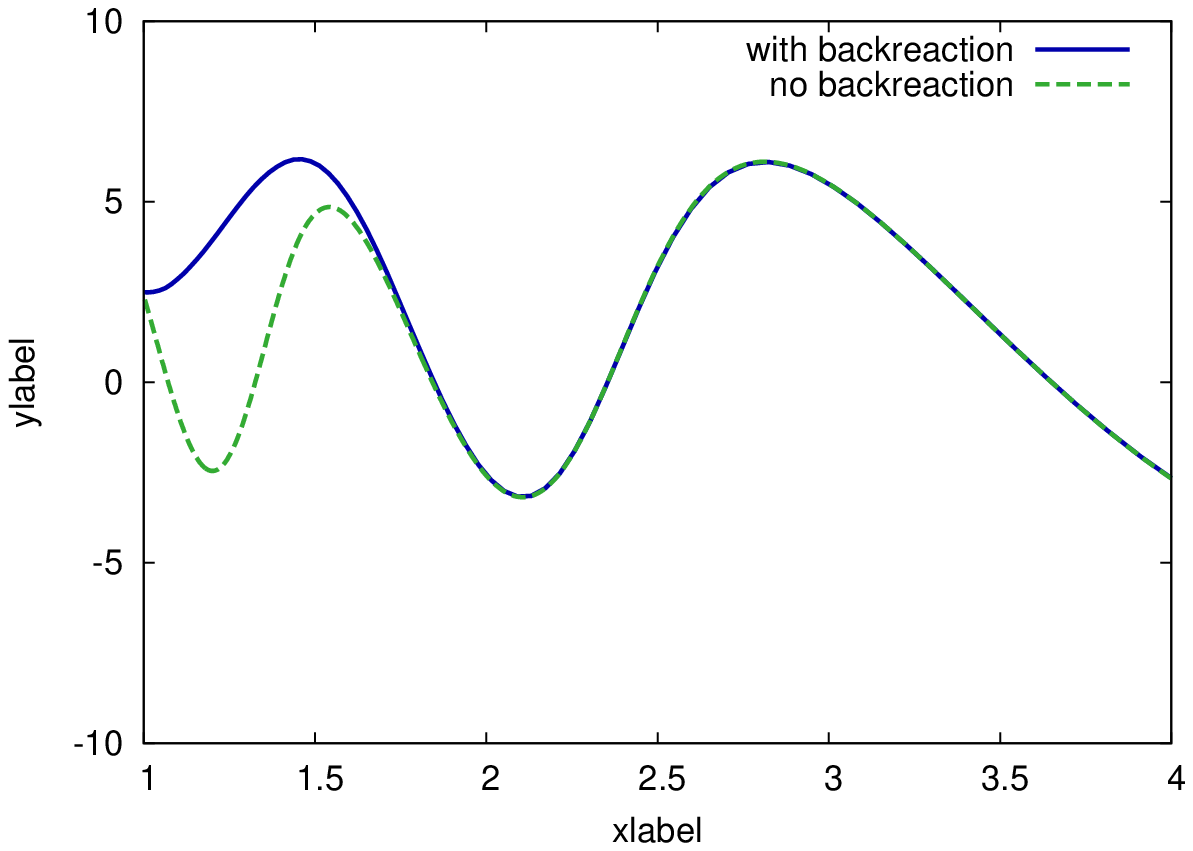}
	\end{center}
	\caption{Evolution of the equation of state of quintessence,
	$w_\varphi$, the background cosmon field $\bar \varphi$, and its
	time derivative $\bar \varphi '$
	(including backreaction effects: blue solid; compared
	to the background equations: green dashed).}
	\label{fig:phi}
\end{figure}

Due to \eq{qphi}, we can understand the evolution of the
deceleration parameter, \fig{deceleration}, by examining the
background evolution of the cosmon. We show the equation of state
$w_\varphi$ as well as the average values $\bar \varphi$ and $\bar
\varphi'$ in \fig{phi}.

In the lowermost plot, we see that once backreaction effects become
important (at redshifts around $z \approx 1.5$), the time derivative
$\bar \varphi'$ takes significantly larger values than it would be
expected neglecting backreaction. This shows that, in fact, the
neutrino fluid is less effective in stopping the cosmon evolution.
This is also reflected in the plot of $\bar \varphi$, which continues
to grow although, neglecting backreaction, a mere oscillation around a
very slowly increasing value has set in at $z \lesssim 1.5$. As a
consequence, the equation of state $w_\varphi$ is, due to the
backreaction, further away from the cosmological constant value $w =
-1$. 

We conclude that the study of non--linear structure formation in
growing neutrino quintessence turns out to be crucial for the
understanding of its correct background evolution as well. A realistic
model with the correct fraction of dark energy today might need a
readjustment of the model parameters.

\subsection{Matter perturbations}
\label{sec:matter}

The dynamics of matter perturbations in our simulation are only
described by the gravitational force $\propto \vec \nabla \Psi$ and the
Hubble damping. The effective matter particles are accelerated
according to the Newtonian law of motion, \eq{eomm}, and we do not
differentiate between baryons and dark matter. Matter is not coupled
to the neutrinos or to the cosmon except via gravity.

The gravitational effects due to the neutrinos and the cosmon include
their gravitational potential and their impact on the expansion
history. In particular, the large neutrino structures investigated in
\sec{neutrino} induce gradients in the gravitational potential leading
to an additional acceleration of matter particles. We thus expect that
the amplitudes of matter perturbations are enhanced compared to the
$\Lambda$CDM case \cite{Baldi11, Ayaita09}.

In order to study this effect, we follow the evolution of the dimensionless matter
spectrum,
\begin{equation}
	\Delta_m^2(k) = \frac{k^3}{2 \pi^2} P_m(k).
	\label{eq:Delta_m}
\end{equation}
The concrete values of $\Delta_m(k)$ of course depend on the model parameters we
have chosen for growing neutrino quintessence, \tab{model}. Hence, in order to
isolate the impact of non--linear neutrino structures on the growth of matter
perturbations, we normalize $\Delta_m(k)$ by the case where matter only feels
its own gravitational potential. This means, we have started another simulation
run following the evolution of only matter particles in a universe with the
expansion history of unperturbed growing neutrino quintessence. Since the
expansion history is, for the chosen set of model parameters, similar to
$\Lambda$CDM, we somewhat imprecisely label the corresponding amplitudes by
$\Delta_m^{\Lambda\text{CDM}}(k)$.

\begin{figure}[htb!]
	\begin{center}
		\psfrag{xlabel}[B][c][.8][0]{scale $k$ [$h$/Mpc] }
		\psfrag{ylabel}[B][c][.8][0]{enhancement of the matter spectrum}
		\includegraphics[width=0.45\textwidth]{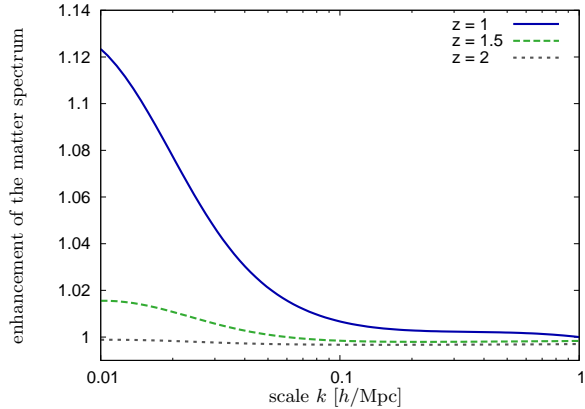}
	\end{center}
	\caption{The evolution of the relative matter spectrum
	$\Delta_m / \Delta_m^{\Lambda\mathrm{CDM}}$.}
	\label{fig:relative_matter_spectra}
\end{figure}

The quotients $\Delta_m(k)/\Delta^{\Lambda\mathrm{CDM}}_m(k)$ at different
redshifts are shown in \fig{relative_matter_spectra}.

At early times, $z>2$, the distribution of the neutrinos is still very smooth
and matter fluctuations grow only due to their own gravitational potential. At
later times, we observe an enhancement of matter fluctuations at large scales,
where the neutrino fluid is clustering, \cf \fig{film}. At $z=1$, the
amplification exceeds $10 \%$ on the largest scales of our simulation,
whereas it is still below the percent level at small scales.     

By means of the continuity equation, the change $\delta_m'$ of the
matter density perturbation is linked to the velocity field $\vec
v^\text{pec}_m$. If the density perturbations show a steep increase, as we
observe it on large scales, large peculiar velocities must be present.

We can measure the matter {\it bulk flow} $\bar {\vec v}^\text{pec}_m$ in a
subvolume $V$ of our simulation defined as the average peculiar
velocity,
\begin{equation}
	\bar {\vec v}^\text{pec}_{m} 
	= \frac{\int_{V}^{} \dd^3 x\, \sqrt{\tilde g}\, \vec v^\text{pec}_m}{\int_{V}^{}
	\dd^3 x\, \sqrt{\tilde g}}.\ \
	\label{eq:bulk}
\end{equation}
Next, we introduce the variance of the bulk flow $U_l^2$ on different comoving scales $l$
as a measure of the expected bulk velocity in volumes of size $l^3$. Technically, we divide the simulation box into $n$ subvolumes $V_i$ of
equal size $l^3 = L^3 / n$. We then obtain the average peculiar velocity
$\bar {\vec v}^\text{pec}_{m,i}$ within the boxes $i$ and compute the variance according to
\begin{equation}
	U_l^2 = \frac{1}{n} \sum_{i}^{} \left( \bar {\vec v}^\text{pec}_{m,i}
	\right)^2.
	\label{eq:ul}
\end{equation}

\begin{figure}[htb!]
	\begin{center}
		\psfrag{xlabel}[B][c][.8][0]{redshift $z$}
		\psfrag{ylabel}[B][c][.8][0]{enhancement of matter bulk
		velocities}
		\includegraphics[width=0.45\textwidth]{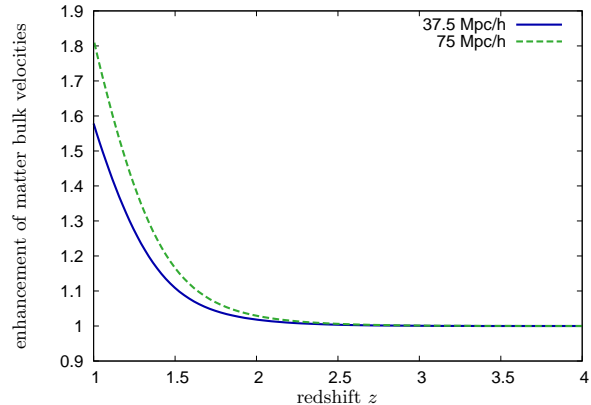}
	\end{center}
	\caption{The enhancement of matter bulk velocities
	$U_l / U_l^{\Lambda\mathrm{CDM}}$ in subboxes of
	volume $l^3$ for $l = 37.5\, h^{-1}$Mpc (blue solid) and $l = 75\,
	h^{-1}$Mpc (green dashed).}
	\label{fig:bulks}
\end{figure}

We show the evolution of $U_l$ for two large scales in \fig{bulks}.

Indeed, the enhancement of bulk flows is much more pronounced than the
increase of density perturbations. This feature of growing neutrino
quintessence has been studied in \cite{Baldi11, Ayaita09}. The
increase of bulk flows reaches factors of about $1.5$ to $2$ at $z = 1$, and it
is growing.

\begin{figure}[htb!]
	\begin{center}
		\psfrag{xlabel}[B][c][.8][0]{redshift $z$}
		\psfrag{ylabel}[B][c][.8][0]{enhancement of the gravitational
		potential}
		\includegraphics[width=0.45\textwidth]{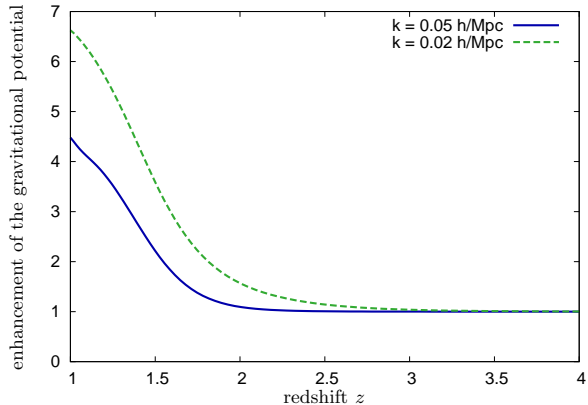}
	\end{center}
	\caption{The evolution of the gravitational potential
	$\Psi(z, k)/\Psi^{\Lambda\text{CDM}}(z, k)$
	for modes $k = 0.05\, h/$Mpc (blue solid) and $k = 0.02\, h/$Mpc
	(green dashed).}
	\label{fig:potev}
\end{figure}

The large bulk flows are the consequence of large--scale gradients in
the gravitational potential $\Psi$. Whereas matter perturbations are
still linear on scales $k \lesssim 0.1\,h/$Mpc, the neutrino density field is highly inhomogeneous on
these and larger scales. It comes as no surprise that the non--linear
neutrino structures dominate the large--scale gravitational potential.
In order to quantify this, we estimate the power spectrum
$P_\Psi(z,k)$ of the gravitational potential. The dimensionless spectrum
\begin{equation}
	\Psi^2(z, k) \equiv \frac{k^3}{2\pi^2}\,P_\Psi(z,k)
	\label{eq:psik}
\end{equation}
is a measure for the fluctuation variance of the cosmological
gravitational potential on spatial volumes $\approx \pi^3 / k^3$. We
compare the large--scale evolution of $\Psi(z, k)$ (including neutrino
structures) with $\Psi^{\Lambda\text{CDM}}(z, k)$ (neglecting neutrino
structure formation) in \fig{potev}. Since $\Phi = \Psi$ on large
scales to a good approximation, \cf \sec{relativistic}, the
corresponding plot for $\Phi(z, k)$ is visually identical.

Once the non--linear effects of neutrino structure formation become
important at $z \lesssim 2$, the large--scale gravitational potential
grows drastically compared to $\Lambda$CDM. Since the matter
perturbations have only grown moderately under the influence of
neutrino structures until $z = 1$, see \fig{relative_matter_spectra}, this effect is
due to the neutrino--induced gravitational potential. Consequently,
\fig{potev} shows that the large--scale potential of the neutrino
structures exceeds that of matter by about an order of magnitude. The
large--scale gravitational potential $\Psi(k)$, \eq{psik}, at $z = 1$
reaches values between $10^{-5}$ and $10^{-4}$ in our simulation.

Whereas the large--scale gravitational potentials in $\Lambda$CDM are
constant during matter domination and very slowly decay thereafter,
growing neutrino quintessence shows a completely different behavior.
During the formation of neutrino structures, the large--scale
gravitational potentials become deeper. This is expected to have a
drastic impact on the integrated Sachs--Wolfe effect, which is
sensitive to the time derivative $(\Psi+\Phi)'$. In this way, the
evolution of the gravitational potentials might be decisive for
scrutinizing the model. A continuing growth of the gravitational
potentials for $z < 1$ would clearly be in conflict with observational
results \cite{Ho08}. The evolution for $z < 1$, however, could be very
different. The virialization of the neutrino structures might stop the
growth of the neutrino--induced gravitational potential.

\section{Summary}
\label{sec:summary}

The cosmological dynamics of growing neutrino quintessence show a
much higher complexity than the standard $\Lambda$CDM cosmology or
models of uncoupled dark energy. This work has made a crucial step
towards a consistent simulation of growing neutrino quintessence
including all its relevant effects.

The standard methods of linear perturbation theory and (Newtonian)
$N$--body simulations, while very successful in the $\Lambda$CDM case,
rely on approximations in conflict with the effects of growing
neutrino quintessence. Linear perturbation theory breaks down on all
scales \cite{Mota08}, and in Newtonian $N$--body simulations, the
particles' velocities exceed the speed of light \cite{Baldi11}.
Furthermore, the local variation of the average neutrino mass due to
perturbations in the dark energy scalar field could so far only be
studied in idealized configurations \cite{Nunes11}. None of the
applied methods has shed light on the expected backreaction
\cite{Schrempp09} of the highly non--linear neutrino perturbations on
the evolution of the cosmological background.

We have thus decided to develop a new method coping with these
challenges right from the start. This effort has led to an $N$--body
based simulation, designed from scratch and adjusted to growing
neutrino quintessence. In particular, we incorporate local variations
in the dark energy scalar field and respect relativistic dynamics for
the neutrino component. We run the simulation for an exemplary choice
of model parameters. Our results show that all the effects that had to
be neglected in previous approaches are indeed important.

The simulation confirms the formation of large--scale neutrino
structures (already predicted in \cite{Brouzakis07}), and we give a
detailed description of their evolution (\sec{growth}).  As a main
result, we observe a significant neutrino mass suppression of an order
of magnitude inside concentrated structures (\sec{individual})
agreeing with \cite{Nunes11}. This is the consequence of significant
inhomogeneities in the dark energy scalar field, the cosmon $\varphi$.
Once these become too large, our method encounters numerical
difficulties (\sec{numerical}), which is why we do not follow the
cosmological evolution for redshifts $z<1$.

We find that the velocities of neutrinos are accelerated to 
relativistic values during the process of structure formation. The
average neutrino equation of state reaches values $w_\nu \approx 0.1$
at $z\approx 1$ (\sec{relativistic}). Further, we have shown that the
relativistic motion induces a difference between the two gravitational
potentials. In the vicinity of large neutrino structures, the
difference $\Phi-\Psi$ typically amounts to $\sim 10^{-5}$.

Local mass variations and relativistic velocities of the neutrinos
lead to strong backreaction effects (\sec{backreaction}). We have
demonstrated that the backreaction can notably modify the late--time
expansion of the Universe (\sec{backreactionandquintessence}). It is
likely that the onset of the accelerated expansion occurs later as
compared to a homogeneous approximation.  

The evolution of matter perturbations is affected by the
neutrino--induced gravitational potential, which dominates on large
scales (\sec{matter}). As a consequence, large--scale bulk flows of
matter are amplified by a factor close to two at $z=1$. The density 
field, however, is much less affected. On large scales, we have found
an effect of about $10 \%$.  \newline

The current stage of the simulation method presented in this work is
already promising, and the efforts to extend its range of
applicability will continue. When the cosmological evolution can be
followed until $z = 0$ for a collection of different model parameters,
a confrontation of growing neutrino quintessence with observational
data becomes possible.

So far, we are only able to identify several effects that might be
interesting regarding observations. The large--scale neutrino--induced
gravitational potentials could be observable directly via
gravitational lensing or indirectly via the resulting large--scale
bulk flows of matter and the enhanced density power spectrum. The time
evolution of the gravitational potential, on the other hand, leaves
imprints in the integrated Sachs--Wolfe effect. In contrast to
$\Lambda$CDM, we have observed increasing large--scale potentials in
growing neutrino quintessence during the formation of neutrino
structures.

The main motivation to search for alternatives to the cosmological
constant scenario are the difficulties to understand the tiny value of
$\Lambda$ and the {\it why now} problem. Any competing model should
avoid them in the first place. Growing neutrino quintessence offers a
mechanism to solve the {\it why now} problem of dark energy. With a
rich phenomenology beyond the standard scenario, the prospects are
promising that the model eventually can be put to stringent tests once
its non--linear evolution is completely understood.  \newline

\begin{acknowledgments}
	We thank David F.\/ Mota, who kindly provided his numerical
	implementation of linear perturbation theory in growing neutrino
	quintessence, Bj\"{o}rn M.\/ Sch\"{a}fer for numerous suggestions
	and valuable ideas, and Luca Amendola for interesting discussions.
	We acknowledge support from the DFG Transregional Collaborative
	Research Centre on the ``Dark Universe''.
\end{acknowledgments}

\bibliography{gnqsim}

\begin{thebibliography}{10}%
\makeatletter
\providecommand \@ifxundefined [1]{%
 \ifx #1\undefined \expandafter \@firstoftwo
 \else \expandafter \@secondoftwo
\fi
}%
\providecommand \@ifnum [1]{%
 \ifnum #1\expandafter \@firstoftwo
 \else \expandafter \@secondoftwo
\fi
}%
\providecommand \enquote [1]{``#1''}%
\providecommand \bibnamefont  [1]{#1}%
\providecommand \bibfnamefont [1]{#1}%
\providecommand \citenamefont [1]{#1}%
\providecommand\href[0]{\@sanitize\@href}%
\providecommand\@href[1]{\endgroup\@@startlink{#1}\endgroup\@@href}%
\providecommand\@@href[1]{#1\@@endlink}%
\providecommand \@sanitize [0]{\begingroup\catcode`\&12\catcode`\#12\relax}%
\@ifxundefined \pdfoutput {\@firstoftwo}{%
 \@ifnum{\z@=\pdfoutput}{\@firstoftwo}{\@secondoftwo}%
}{%
 \providecommand\@@startlink[1]{\leavevmode\special{html:<a href="#1">}}%
 \providecommand\@@endlink[0]{\special{html:</a>}}%
}{%
 \providecommand\@@startlink[1]{%
  \leavevmode
  \pdfstartlink
   attr{/Border[0 0 1 ]/H/I/C[0 1 1]}%
   user{/Subtype/Link/A<</Type/Action/S/URI/URI(#1)>>}%
  \relax
 }%
 \providecommand\@@endlink[0]{\pdfendlink}%
}%
\providecommand \url  [0]{\begingroup\@sanitize \@url }%
\providecommand \@url [1]{\endgroup\@href {#1}{\urlprefix}}%
\providecommand \urlprefix [0]{URL }%
\providecommand \Eprint[0]{\href }%
\@ifxundefined \urlstyle {%
  \providecommand \doi [1]{doi:\discretionary{}{}{}#1}%
}{%
  \providecommand \doi [0]{doi:\discretionary{}{}{}\begingroup
  \urlstyle{rm}\Url }%
}%
\providecommand \doibase [0]{http://dx.doi.org/}%
\providecommand \Doi[1]{\href{\doibase#1}}%
\providecommand \bibAnnote [3]{%
  \BibitemShut{#1}%
  \begin{quotation}\noindent
    \textsc{Key:}\ #2\\\textsc{Annotation:}\ #3%
  \end{quotation}%
}%
\providecommand \bibAnnoteFile [2]{%
  \IfFileExists{#2}{\bibAnnote {#1} {#2} {\input{#2}}}{}%
}%
\providecommand \typeout [0]{\immediate \write \m@ne }%
\providecommand \selectlanguage [0]{\@gobble}%
\providecommand \bibinfo [0]{\@secondoftwo}%
\providecommand \bibfield [0]{\@secondoftwo}%
\providecommand \translation [1]{[#1]}%
\providecommand \BibitemOpen[0]{}%
\providecommand \bibitemStop [0]{}%
\providecommand \bibitemNoStop [0]{.\EOS\space}%
\providecommand \EOS [0]{\spacefactor3000\relax}%
\providecommand \BibitemShut [1]{\csname bibitem#1\endcsname}%
\bibitem{WoodVasey07}%
  \BibitemOpen
  \bibfield{author}{%
  \bibinfo {author} {\bibfnamefont{W.~M.}\ \bibnamefont{Wood-Vasey}}
  \emph{et~al.} (\bibinfo {collaboration} {ESSENCE}),\ }%
  \bibfield{journal}{%
  \Doi{10.1086/518642}{\bibinfo {journal} {Astrophys. J.}}\ }%
  \textbf{\bibinfo {volume} {666}},\ \bibinfo {pages} {694} (\bibinfo {year}
  {2007}),\
  \Eprint{http://arxiv.org/abs/astro-ph/0701041}{arXiv:astro-ph/0701041}%
  \bibAnnoteFile{NoStop}{WoodVasey07}%
\bibitem{Reid09}%
  \BibitemOpen
  \bibfield{author}{%
  \bibinfo {author} {\bibfnamefont{B.~A.}\ \bibnamefont{Reid}} \emph{et~al.},\
  }%
  \bibfield{journal}{%
  \Doi{10.1111/j.1365-2966.2010.16276.x}{\bibinfo {journal} {Mon. Not. Roy.
  Astron. Soc.}}\ }%
  \textbf{\bibinfo {volume} {404}},\ \bibinfo {pages} {60} (\bibinfo {year}
  {2010}),\ \Eprint{http://arxiv.org/abs/0907.1659}{arXiv:0907.1659
  [astro-ph.CO]}%
  \bibAnnoteFile{NoStop}{Reid09}%
\bibitem{Komatsu10}%
  \BibitemOpen
  \bibfield{author}{%
  \bibinfo {author} {\bibfnamefont{E.}~\bibnamefont{Komatsu}} \emph{et~al.}
  (\bibinfo {collaboration} {WMAP}),\ }%
  \bibfield{journal}{%
  \Doi{10.1088/0067-0049/192/2/18}{\bibinfo {journal} {Astrophys. J. Suppl.}}\
  }%
  \textbf{\bibinfo {volume} {192}},\ \bibinfo {pages} {18} (\bibinfo {year}
  {2011}),\ \Eprint{http://arxiv.org/abs/1001.4538}{arXiv:1001.4538
  [astro-ph.CO]}%
  \bibAnnoteFile{NoStop}{Komatsu10}%
\bibitem{Sullivan11}%
  \BibitemOpen
  \bibfield{author}{%
  \bibinfo {author} {\bibfnamefont{M.}~\bibnamefont{Sullivan}} \emph{et~al.},\
  }%
  \bibfield{journal}{%
  \Doi{10.1088/0004-637X/737/2/102}{\bibinfo {journal} {Astrophys. J.}}\ }%
  \textbf{\bibinfo {volume} {737}},\ \bibinfo {pages} {102} (\bibinfo {year}
  {2011}),\ \Eprint{http://arxiv.org/abs/1104.1444}{arXiv:1104.1444
  [astro-ph.CO]}%
  \bibAnnoteFile{NoStop}{Sullivan11}%
\bibitem{Amendola07}%
  \BibitemOpen
  \bibfield{author}{%
  \bibinfo {author} {\bibfnamefont{L.}~\bibnamefont{Amendola}}, \bibinfo
  {author} {\bibfnamefont{M.}~\bibnamefont{Baldi}},\ and\ \bibinfo {author}
  {\bibfnamefont{C.}~\bibnamefont{Wetterich}},\ }%
  \bibfield{journal}{%
  \Doi{10.1103/PhysRevD.78.023015}{\bibinfo {journal} {Phys. Rev.}}\ }%
  \textbf{\bibinfo {volume} {D78}},\ \bibinfo {pages} {023015} (\bibinfo {year}
  {2008}),\ \Eprint{http://arxiv.org/abs/0706.3064}{arXiv:0706.3064
  [astro-ph]}%
  \bibAnnoteFile{NoStop}{Amendola07}%
\bibitem{Wetterich07}%
  \BibitemOpen
  \bibfield{author}{%
  \bibinfo {author} {\bibfnamefont{C.}~\bibnamefont{Wetterich}},\ }%
  \bibfield{journal}{%
  \Doi{10.1016/j.physletb.2007.08.060}{\bibinfo {journal} {Phys. Lett.}}\ }%
  \textbf{\bibinfo {volume} {B655}},\ \bibinfo {pages} {201} (\bibinfo {year}
  {2007}),\ \Eprint{http://arxiv.org/abs/0706.4427}{arXiv:0706.4427 [hep-ph]}%
  \bibAnnoteFile{NoStop}{Wetterich07}%
\bibitem{Mota08}%
  \BibitemOpen
  \bibfield{author}{%
  \bibinfo {author} {\bibfnamefont{D.~F.}\ \bibnamefont{Mota}}, \bibinfo
  {author} {\bibfnamefont{V.}~\bibnamefont{Pettorino}}, \bibinfo {author}
  {\bibfnamefont{G.}~\bibnamefont{Robbers}},\ and\ \bibinfo {author}
  {\bibfnamefont{C.}~\bibnamefont{Wetterich}},\ }%
  \bibfield{journal}{%
  \Doi{10.1016/j.physletb.2008.03.060}{\bibinfo {journal} {Phys. Lett.}}\ }%
  \textbf{\bibinfo {volume} {B663}},\ \bibinfo {pages} {160} (\bibinfo {year}
  {2008}),\ \Eprint{http://arxiv.org/abs/0802.1515}{arXiv:0802.1515
  [astro-ph]}%
  \bibAnnoteFile{NoStop}{Mota08}%
\bibitem{Wintergerst09}%
  \BibitemOpen
  \bibfield{author}{%
  \bibinfo {author} {\bibfnamefont{N.}~\bibnamefont{Wintergerst}}, \bibinfo
  {author} {\bibfnamefont{V.}~\bibnamefont{Pettorino}}, \bibinfo {author}
  {\bibfnamefont{D.~F.}\ \bibnamefont{Mota}},\ and\ \bibinfo {author}
  {\bibfnamefont{C.}~\bibnamefont{Wetterich}},\ }%
  \bibfield{journal}{%
  \Doi{10.1103/PhysRevD.81.063525}{\bibinfo {journal} {Phys. Rev.}}\ }%
  \textbf{\bibinfo {volume} {D81}},\ \bibinfo {pages} {063525} (\bibinfo {year}
  {2010}),\ \Eprint{http://arxiv.org/abs/0910.4985}{arXiv:0910.4985
  [astro-ph.CO]}%
  \bibAnnoteFile{NoStop}{Wintergerst09}%
\bibitem{Baldi11}%
  \BibitemOpen
  \bibfield{author}{%
  \bibinfo {author} {\bibfnamefont{M.}~\bibnamefont{Baldi}}, \bibinfo {author}
  {\bibfnamefont{V.}~\bibnamefont{Pettorino}}, \bibinfo {author}
  {\bibfnamefont{L.}~\bibnamefont{Amendola}},\ and\ \bibinfo {author}
  {\bibfnamefont{C.}~\bibnamefont{Wetterich}}}%
   (\bibinfo {year} {2011}),\
  \Eprint{http://arxiv.org/abs/1106.2161}{arXiv:1106.2161 [astro-ph.CO]}%
  \bibAnnoteFile{NoStop}{Baldi11}%
\bibitem{Nunes11}%
  \BibitemOpen
  \bibfield{author}{%
  \bibinfo {author} {\bibfnamefont{N.~J.}\ \bibnamefont{Nunes}}, \bibinfo
  {author} {\bibfnamefont{L.}~\bibnamefont{Schrempp}},\ and\ \bibinfo {author}
  {\bibfnamefont{C.}~\bibnamefont{Wetterich}},\ }%
  \bibfield{journal}{%
  \Doi{10.1103/PhysRevD.83.083523}{\bibinfo {journal} {Phys. Rev.}}\ }%
  \textbf{\bibinfo {volume} {D83}},\ \bibinfo {pages} {083523} (\bibinfo {year}
  {2011}),\ \Eprint{http://arxiv.org/abs/1102.1664}{arXiv:1102.1664
  [astro-ph.CO]}%
  \bibAnnoteFile{NoStop}{Nunes11}%
\bibitem{Wetterich94}%
  \BibitemOpen
  \bibfield{author}{%
  \bibinfo {author} {\bibfnamefont{C.}~\bibnamefont{Wetterich}},\ }%
  \bibfield{journal}{%
  \bibinfo {journal} {Astron. Astrophys.}\ }%
  \textbf{\bibinfo {volume} {301}},\ \bibinfo {pages} {321} (\bibinfo {year}
  {1995}),\ \Eprint{http://arxiv.org/abs/hep-th/9408025}{arXiv:hep-th/9408025}%
  \bibAnnoteFile{NoStop}{Wetterich94}%
\bibitem{Amendola99}%
  \BibitemOpen
  \bibfield{author}{%
  \bibinfo {author} {\bibfnamefont{L.}~\bibnamefont{Amendola}},\ }%
  \bibfield{journal}{%
  \Doi{10.1103/PhysRevD.62.043511}{\bibinfo {journal} {Phys. Rev.}}\ }%
  \textbf{\bibinfo {volume} {D62}},\ \bibinfo {pages} {043511} (\bibinfo {year}
  {2000}),\
  \Eprint{http://arxiv.org/abs/astro-ph/9908023}{arXiv:astro-ph/9908023}%
  \bibAnnoteFile{NoStop}{Amendola99}%
\bibitem{Doran07}%
  \BibitemOpen
  \bibfield{author}{%
  \bibinfo {author} {\bibfnamefont{M.}~\bibnamefont{Doran}}, \bibinfo {author}
  {\bibfnamefont{G.}~\bibnamefont{Robbers}},\ and\ \bibinfo {author}
  {\bibfnamefont{C.}~\bibnamefont{Wetterich}},\ }%
  \bibfield{journal}{%
  \Doi{10.1103/PhysRevD.75.023003}{\bibinfo {journal} {Phys. Rev.}}\ }%
  \textbf{\bibinfo {volume} {D75}},\ \bibinfo {pages} {023003} (\bibinfo {year}
  {2007}),\
  \Eprint{http://arxiv.org/abs/astro-ph/0609814}{arXiv:astro-ph/0609814}%
  \bibAnnoteFile{NoStop}{Doran07}%
\bibitem{Reichardt11}%
  \BibitemOpen
  \bibfield{author}{%
  \bibinfo {author} {\bibfnamefont{C.~L.}\ \bibnamefont{Reichardt}}, \bibinfo
  {author} {\bibfnamefont{R.}~\bibnamefont{de~Putter}}, \bibinfo {author}
  {\bibfnamefont{O.}~\bibnamefont{Zahn}},\ and\ \bibinfo {author}
  {\bibfnamefont{Z.}~\bibnamefont{Hou}},\ }%
  \bibfield{journal}{%
  \bibinfo {journal} {Astrophys.J.}\ }%
  \textbf{\bibinfo {volume} {749}},\ \bibinfo {pages} {L9} (\bibinfo {year}
  {2012}),\ \Eprint{http://arxiv.org/abs/1110.5328}{arXiv:1110.5328
  [astro-ph.CO]}%
  \bibAnnoteFile{NoStop}{Reichardt11}%
\bibitem{Brill57}%
  \BibitemOpen
  \bibfield{author}{%
  \bibinfo {author} {\bibfnamefont{D.~R.}\ \bibnamefont{Brill}}\ and\ \bibinfo
  {author} {\bibfnamefont{J.~A.}\ \bibnamefont{Wheeler}},\ }%
  \bibfield{journal}{%
  \Doi{10.1103/RevModPhys.29.465}{\bibinfo {journal} {Rev. Mod. Phys.}}\ }%
  \textbf{\bibinfo {volume} {29}},\ \bibinfo {pages} {465} (\bibinfo {year}
  {1957})%
  \bibAnnoteFile{NoStop}{Brill57}%
\bibitem{Ma95}%
  \BibitemOpen
  \bibfield{author}{%
  \bibinfo {author} {\bibfnamefont{C.-P.}\ \bibnamefont{Ma}}\ and\ \bibinfo
  {author} {\bibfnamefont{E.}~\bibnamefont{Bertschinger}},\ }%
  \bibfield{journal}{%
  \Doi{10.1086/176550}{\bibinfo {journal} {Astrophys. J.}}\ }%
  \textbf{\bibinfo {volume} {455}},\ \bibinfo {pages} {7} (\bibinfo {year}
  {1995}),\
  \Eprint{http://arxiv.org/abs/astro-ph/9506072}{arXiv:astro-ph/9506072}%
  \bibAnnoteFile{NoStop}{Ma95}%
\bibitem{Brouzakis07}%
  \BibitemOpen
  \bibfield{author}{%
  \bibinfo {author} {\bibfnamefont{N.}~\bibnamefont{Brouzakis}}, \bibinfo
  {author} {\bibfnamefont{N.}~\bibnamefont{Tetradis}},\ and\ \bibinfo {author}
  {\bibfnamefont{C.}~\bibnamefont{Wetterich}},\ }%
  \bibfield{journal}{%
  \Doi{10.1016/j.physletb.2008.05.068}{\bibinfo {journal} {Phys. Lett.}}\ }%
  \textbf{\bibinfo {volume} {B665}},\ \bibinfo {pages} {131} (\bibinfo {year}
  {2008}),\ \Eprint{http://arxiv.org/abs/0711.2226}{arXiv:0711.2226
  [astro-ph]}%
  \bibAnnoteFile{NoStop}{Brouzakis07}%
\bibitem{Wetterich01}%
  \BibitemOpen
  \bibfield{author}{%
  \bibinfo {author} {\bibfnamefont{C.}~\bibnamefont{Wetterich}},\ }%
  \bibfield{journal}{%
  \Doi{10.1103/PhysRevD.67.043513}{\bibinfo {journal} {Phys. Rev.}}\ }%
  \textbf{\bibinfo {volume} {D67}},\ \bibinfo {pages} {043513} (\bibinfo {year}
  {2003}),\
  \Eprint{http://arxiv.org/abs/astro-ph/0111166}{arXiv:astro-ph/0111166}%
  \bibAnnoteFile{NoStop}{Wetterich01}%
\bibitem{Schrempp09}%
  \BibitemOpen
  \bibfield{author}{%
  \bibinfo {author} {\bibfnamefont{L.}~\bibnamefont{Schrempp}}\ and\ \bibinfo
  {author} {\bibfnamefont{I.}~\bibnamefont{Brown}},\ }%
  \bibfield{journal}{%
  \Doi{10.1088/1475-7516/2010/05/023}{\bibinfo {journal} {JCAP}}\ }%
  \textbf{\bibinfo {volume} {1005}},\ \bibinfo {pages} {023} (\bibinfo {year}
  {2010}),\ \Eprint{http://arxiv.org/abs/0912.3157}{arXiv:0912.3157
  [astro-ph.CO]}%
  \bibAnnoteFile{NoStop}{Schrempp09}%
\bibitem{Efstathiou85}%
  \BibitemOpen
  \bibfield{author}{%
  \bibinfo {author} {\bibfnamefont{G.}~\bibnamefont{Efstathiou}}, \bibinfo
  {author} {\bibfnamefont{M.}~\bibnamefont{Davis}}, \bibinfo {author}
  {\bibfnamefont{C.~S.}\ \bibnamefont{Frenk}},\ and\ \bibinfo {author}
  {\bibfnamefont{S.~D.~M.}\ \bibnamefont{White}},\ }%
  \bibfield{journal}{%
  \Doi{10.1086/191003}{\bibinfo {journal} {Astrophys. J. Suppl.}}\ }%
  \textbf{\bibinfo {volume} {57}},\ \bibinfo {pages} {241} (\bibinfo {year}
  {1985})%
  \bibAnnoteFile{NoStop}{Efstathiou85}%
\bibitem{Dolag08}%
  \BibitemOpen
  \bibfield{author}{%
  \bibinfo {author} {\bibfnamefont{K.}~\bibnamefont{Dolag}}, \bibinfo {author}
  {\bibfnamefont{S.}~\bibnamefont{Borgani}}, \bibinfo {author}
  {\bibfnamefont{S.}~\bibnamefont{Schindler}}, \bibinfo {author}
  {\bibfnamefont{A.}~\bibnamefont{Diaferio}},\ and\ \bibinfo {author}
  {\bibfnamefont{A.~M.}\ \bibnamefont{Bykov}}}%
   (\bibinfo {year} {2008}),\
  \Eprint{http://arxiv.org/abs/0801.1023}{arXiv:0801.1023 [astro-ph]}%
  \bibAnnoteFile{NoStop}{Dolag08}%
\bibitem{Springel05}%
  \BibitemOpen
  \bibfield{author}{%
  \bibinfo {author} {\bibfnamefont{V.}~\bibnamefont{Springel}},\ }%
  \bibfield{journal}{%
  \Doi{10.1111/j.1365-2966.2005.09655.x}{\bibinfo {journal}
  {Mon.Not.Roy.Astron.Soc.}}\ }%
  \textbf{\bibinfo {volume} {364}},\ \bibinfo {pages} {1105} (\bibinfo {year}
  {2005}),\
  \Eprint{http://arxiv.org/abs/astro-ph/0505010}{arXiv:astro-ph/0505010
  [astro-ph]}%
  \bibAnnoteFile{NoStop}{Springel05}%
\bibitem{Ayaita09}%
  \BibitemOpen
  \bibfield{author}{%
  \bibinfo {author} {\bibfnamefont{Y.}~\bibnamefont{Ayaita}}, \bibinfo {author}
  {\bibfnamefont{M.}~\bibnamefont{Weber}},\ and\ \bibinfo {author}
  {\bibfnamefont{C.}~\bibnamefont{Wetterich}}}%
   (\bibinfo {year} {2009}),\
  \Eprint{http://arxiv.org/abs/0908.2903}{arXiv:0908.2903 [astro-ph.CO]}%
  \bibAnnoteFile{NoStop}{Ayaita09}%
\bibitem{Ho08}%
  \BibitemOpen
  \bibfield{author}{%
  \bibinfo {author} {\bibfnamefont{S.}~\bibnamefont{Ho}}, \bibinfo {author}
  {\bibfnamefont{C.}~\bibnamefont{Hirata}}, \bibinfo {author}
  {\bibfnamefont{N.}~\bibnamefont{Padmanabhan}}, \bibinfo {author}
  {\bibfnamefont{U.}~\bibnamefont{Seljak}},\ and\ \bibinfo {author}
  {\bibfnamefont{N.}~\bibnamefont{Bahcall}},\ }%
  \bibfield{journal}{%
  \Doi{10.1103/PhysRevD.78.043519}{\bibinfo {journal} {Phys.Rev.}}\ }%
  \textbf{\bibinfo {volume} {D78}},\ \bibinfo {pages} {043519} (\bibinfo {year}
  {2008}),\ \Eprint{http://arxiv.org/abs/0801.0642}{arXiv:0801.0642
  [astro-ph]}%
  \bibAnnoteFile{NoStop}{Ho08}%
\end{thebibliography}%

\end{document}